\documentstyle[epsfig,11pt]{article}

\begin{document}

\title{PHOTON EXCHANGE IN NUCLEUS-NUCLEUS COLLISIONS
}

\author{
CARLOS A. BERTULANI\footnote{e-mail address:
bertulani@nscl.msu.edu} \\
NSCL,
Michigan State University\\
East Lansing, MI 48824-1321, USA }

\maketitle

\begin{abstract}
The strong electromagnetic fields in peripheral heavy ion
collisions give rise to photon-photon and photon-nucleus
interactions. I present a general survey of the photon-photon and
photon-hadron physics accessible in these collisions. Among these
processes I discuss the nuclear fragmentation through the
excitation of giant resonances, the Coulomb dissociation method
for application in nuclear astrophysics, and the production of
particles.

\end{abstract}

\section{Peripheral Heavy Ion Collisions}

The field of peripheral atomic collisions was born in 1924, when
E. Fermi had the ingenious idea of relating the atomic processes
induced by fast charged particles to processes induced by
electromagnetic waves. In 1934-1935, Weizs\"{a}cker and Williams
modified Fermi's calculation by including the appropriate
relativistic corrections. The original Fermi's idea is now known
as the Weizs\"{a}cker-Williams method \cite{Fe24,WW34}, an
approximation widely used in coherent processes in atomic, nuclear
and particle physics. In Fermi's method  the electromagnetic
(strong) field generated by a fast particle is replaced by a flux
of equivalent photons (flux mesons or gluons).

\begin{figure}[t]
\centerline{\psfig{file=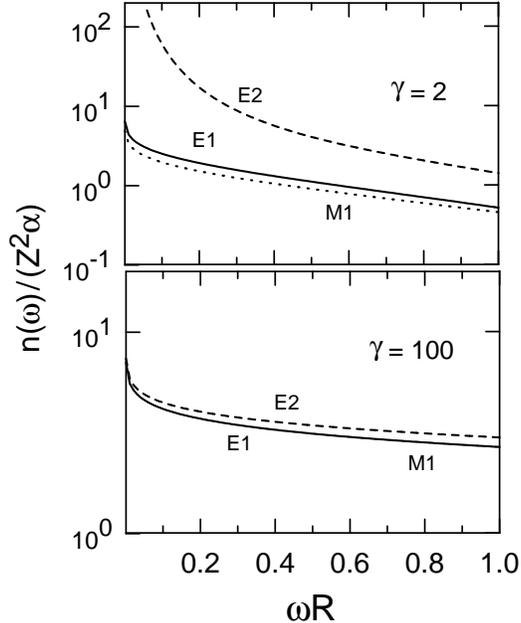,width=7cm}} \vspace*{8pt}
\caption{Virtual photon numbers (in units of $Z^2\alpha$) for the
E1, E2 and M1 multipolarities. Upper (lower) figure is for
$\gamma=2$ ($\gamma=100$). $\omega$ is the photon energy and $R$
is the sum of the nuclear radii.} \label{EPA1}
\end{figure}

In peripheral heavy ion collisions (PHIC) the number of equivalent
photons, $n(\omega )$ of  energy $\omega $ can be calculated
classically, or quantum-mechanically. For the electric dipole (E1)
multipolarity both results are identical under the assumption of
very forward scattering \cite{BB85}. In ref. \cite{BB85} the
number of equivalent photons for all multipolarities was
calculated exactly. It was shown that for the electric dipole
multipolarity, E1, the equivalent photon number, $n_{E1}(\omega
)$, coincides with the one deduced by Weizs\"{a}cker and Williams.
It was also shown that in the extreme relativistic collisions the
equivalent photon numbers for all multipolarities agree, i.e,
$n_{E1}(\omega )\sim n_{E2}(\omega )\sim n_{M1}(\omega )\sim ...$.
This is shown in figure \ref{EPA1}.

\begin{figure}[t]
\centerline{\psfig{file=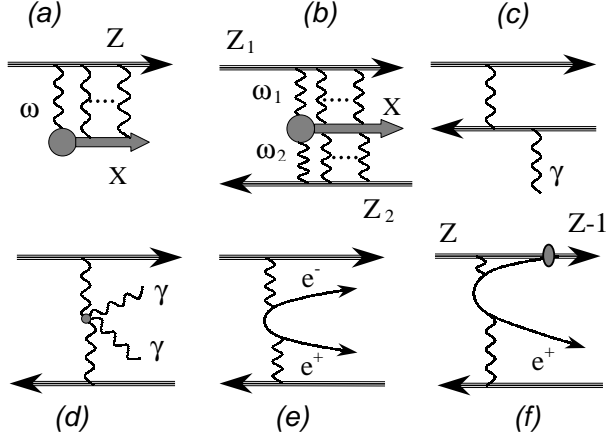,width=8cm}} \vspace*{8pt}
\caption{PHIC processes: (a) one-photon, (b) two-photon, (c)
Bremsstrahlung, (d) Delbr\"uck scattering, (e) pair-production,
and (f) pair-production with capture.} \label{graphs5}
\end{figure}

According to Fermi's idea, the cross sections for one- and
two-photon processes depicted in figure \ref{graphs5}(a,b) are
given by

\begin{equation}
\sigma _{X}=\int d\omega \frac{n\left( \omega \right) }{\omega }\sigma
_{X}^{\gamma }\left( \omega \right) \;,\;{\rm and}\;\;\sigma _{X}=\int
d\omega _{1}d\omega _{2}\frac{n\left( \omega _{1}\right) }{\omega _{1}}\frac{%
n\left( \omega _{2}\right) }{\omega _{2}}\sigma _{X}^{\gamma \gamma }\left(
\omega _{1},\omega _{2}\right) \;,  \label{epa}
\end{equation}
where $\sigma _{X}^{\gamma }\left( \omega \right) $ is the photon-induced
cross section for the energy $\omega $, and $\sigma _{X}^{\gamma \gamma
}\left( \omega _{1},\omega _{2}\right) $ is the two-photon cross section.
Note that we do not refer to the photon momenta. The virtual photons are
real: $q^{2}=0$, a relation always valid for PHIC.

At relativistic bombarding energies one has (in natural units)
\begin{equation}
n(\omega)= {2\over \pi} Z_p^2 \alpha \ln \left( {\gamma \over
\omega R} \right)\ ,
\end{equation}
where $\gamma=(1-v^2)^{-1/2}$ is the relativistic Lorentz factor
in the frame of reference of the target, $Z_pe$ is the projectile
charge and $\alpha=1/137$.

Eq. \ref{epa} is called the equivalent photon approximation (EPA),
or Weizs\"{a}cker-Williams method. It is valid for processes {\it
c, d, e}, and {\it f} in figure \ref{graphs5}. But the
approximation is not applicable to processes {\it a} and {\it b}
in figure \ref{graphs5}, in which multiple virtual photons are
exchanged between the projectile, the target, and/or with the
produced particle(s).

For one-photon processes, e.g., Coulomb fragmentation, $\sigma
_{X}^{\gamma }\left( \omega \right) $ is localized in a small
energy interval and one gets a cross section in the form $\sigma
=A\ln \gamma +B$, where $A$ and $B$ are coefficients depending on
the system. The $\ln\gamma $ factor is due to the equivalent
photon number, $n\left( \omega \right) $, which is approximately
independent of $\omega $ in the integral range of interest. For
two-photon processes, besides the $\ln^{2}\gamma_{c}$ from $n_{1}$
and $n_{2}$, a third $\ln\gamma_{c}$ arises from the integral over
$\omega _{1}$ ($\omega _{1}$ and $\omega _{2}$ are related by
energy conservation). Note that here we used $\gamma _{c}$ of a
heavy ion collider, so that $\gamma =2\gamma _{c}^{2}-1$, with
$\gamma _{c}$ the collider Lorentz gamma factor (e.g., $\gamma
_{c}\sim 100$ for the RHIC collider at Brookhaven).

Most applications of PHIC were reviewed in ref. \cite{BB88}. Since
then a great amount of work has been performed in this field. The
coherent $\gamma$-$\gamma$ and $\gamma$-A interactions in very
peripheral collisions at relativistic ion colliders has been
reviewed recently in ref. \cite{BHTS02}. In this review article I
will present a general account of the theory and the latest
developments in the field of PHIC with fixed targets at
intermediate energies ($E_{Lab} \simeq 100$ MeV/nucleon), and at
ultra-relativistic energies of present day (the Relativistic Heavy
Ion Collider (RHIC), at Brookhaven) and future heavy ion collider
(the Large Hadron Collider (LHC), at CERN) energies .

\section{Relativistic Coulomb Excitation and Fragmentation}

Relativistic Coulomb excitation is now a popular tool for the
investigation of the intrinsic nuclear dynamics and structure of
the colliding nuclei, specially important in reactions involving
radioactive nuclear beams
\cite{BB94,BCH93,Glas98,Au98,BP99,AB95,HJJ95}. The advantage is
that the Coulomb interaction is very well known. The disadvantage
is that the contribution of the nuclear-induced processes also
play a role in some situations. The treatment of the dissociation
problem by nuclear forces is very model dependent, based on
eikonal or multiple Glauber scattering approaches
\cite{BCH93,BBK92,Ie93,BBe93,HRB91}. Among the uncertainties are
the in-medium nucleon-nucleon cross sections, the multiple
scattering process and the separation of stripping from elastic
dissociation of the nuclei \cite{HRB91}. Nonetheless, specially
for the very weakly-bound nuclei, relativistic Coulomb excitation
has lead to very exciting new results
\cite{BCH93,Glas98,BBK92,Ie93,BBe93}.

\subsection{Reactions with Radioactive Beams}
Coulomb breakup of weakly-bound nuclei may involve single or
multiple photon-exchange between the projectile and the target. In
the first case, perturbation theory gives a direct relation
between the data and the matrix elements of electromagnetic
transitions. Such matrix elements are the clearest probes of the
nuclear structure of these nuclei, since one cannot perform
experiments with real photons or with electron scattering off
nuclei far from the stability valley. In the second case, often
called by post-acceleration, or reacceleration, effects
\cite{BBK92,Ie93,BBe93,EBB95,KYS96}, one has to perform a
non-perturbative treatment of the reaction what complicates the
extraction of the electromagnetic (mainly E1) matrix elements.

Among several possible studies of interest, one expects to learn
if the Coulomb-induced breakup proceeds via a resonance or by the
direct dissociation into continuum states \cite{BBK92,Ie93,BBe93}.
There is a strong ongoing effort to use the relativistic Coulomb
excitation technique also for studying bound excited states in
exotic nuclei, to obtain information on gamma-decay widths,
angular momentum, parity, and other properties of hitherto unknown
states \cite{Glas98,AB95,HJJ95}.

\begin{figure}[t]
\centerline{\psfig{file=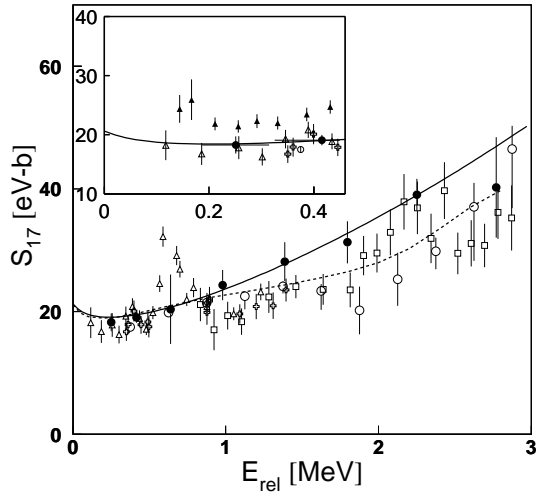,width=7cm}} \vspace*{8pt}
\caption{S-factors ($S_{17}$) for the $^7Be(p,\gamma)^8B$
reaction. The inset is a blowup of the low energy region. See text
for details.} \label{fig4}
\end{figure}

Another application of peripheral collisions with radioactive
nuclear beams is in astrophysics. Radiative capture reactions are
known to play a major role in astrophysical sites, e.g., in a
pre-supernova \cite{Clayton,Rolfs}. Some reactions of interest for
astrophysics, e.g., $^{7}Be\left( p,\gamma \right) ^{8}B$, can be
studied via the inverse photo-dissociation reaction $^{8}B\left(
\gamma ,p\right) ^{7}Be$  \cite{BBH86}. Since the equivalent
photon numbers in eq. \ref{epa} can be calculated theoretically ,
an experimental measurement of the Coulomb breakup reaction
$a+A\longrightarrow b+c+A$ is useful to obtain the corresponding
$\gamma $ induced cross section $\gamma +a\longrightarrow b+c$.
Using detailed balance, this cross section can be related to the
radiative capture cross section $b+c\longrightarrow a+\gamma ,$ of
astrophysical interest \cite{BBH86} (see eq. \ref{Sfactor}).

The radiative capture cross sections in nuclear astrophysics are
often written in terms of the astrophysical S-factor
\cite{Clayton,Rolfs}, defined by
\begin{equation}
S(E)=E\sigma \left(
E\right) \exp \left[ 2\pi \eta \left( E\right) \right] \ ,
\end{equation}
where $\eta \left( E\right) =Z_{b}Z_{c}e^{2}/\hbar \sqrt{2\mu
_{bc}E}$, and $E$ is the relative kinetic energy of the nuclei $b$
and $c$ in the reaction $b(c,\gamma)a$. The cross section for the
dissociation process $a+A\longrightarrow b+c+A$  can be written as

\begin{equation}
{d\sigma \over dE} ={\mu_{bc} c^2\over E_x^3}\
n(E_x)\ {(2J_b+1)(2J_c+1)\over (2J_a+1)} \
S(E) \  \exp \left[ -2\pi \eta \left( E\right) \right]\ ,
\label{Sfactor}
\end{equation}
where $E_x=E+B$ is the sum of the relative energy $E$ and the
binding energy $B$ of the two fragments. $J_i$ is the ground-state
angular momentum of the nucleus i.

\small
\begin{table}
\caption{Radiative capture reactions of interest in nuclear
astrophysics which can be studied with the Coulomb dissociation
method.}
\begin{center}
\begin{tabular}{|l|l|l|}
\hline Reaction & $T_{1/2}$ (projectile) & Astrophysical
application
\\ \hline \hline $^{3}He\left( \alpha ,\gamma \right) ^{7}Be$ &
53.3 days & Solar-neutrinos
\\ \hline
$^{7}Be\left( p,\gamma \right) ^{8}B$ & 770 ms & $^{3}He$
abundance
 \\ \hline
$^{7}Be\left( \alpha ,\gamma \right) ^{11}C$ & 20.4 min &    \\
\hline $^{4}He\left( d,\gamma \right) ^{6}Li$ & Stable &
Primordial nucleosynthesis
\\ \hline
$^{6}Li\left( p,\gamma \right) ^{7}Be$ & 53.3 days &    \\ \hline
$^{6}Li\left( \alpha ,\gamma \right) ^{10}B$ & Stable &
\\ \hline
$^{4}He\left( t,\gamma \right) ^{7}Li$ & Stable &    \\ \hline
$^{7}Li\left( \alpha ,\gamma \right) ^{11}B$ & Stable &    \\
\hline $^{11}B\left( p,\gamma \right) ^{12}C$ & Stable &    \\
\hline $^{9}Be\left( p,\gamma \right) ^{10}B$ & Stable &   \\
\hline $^{10}B\left( p,\gamma \right) ^{11}C$ & 20.4 min &    \\
\hline $^{7}Li\left( n,\gamma \right) ^{8}Li$ & 842 ms &
Inhomogeneous Big Bang  \\ \hline $^{8}Li\left( n,\gamma \right)
^{9}Li$ & 178 ms &   \\ \hline $^{12}C\left( n,\gamma \right)
^{13}C$ & Stable &    \\ \hline $^{14}C\left( n,\gamma \right)
^{15}C$ & 2.45 s &    \\ \hline $^{14}C\left( \alpha ,\gamma
\right) ^{18}O$ & Stable &   \\ \hline
$^{12}C\left( p ,\gamma \right) ^{13}N$ & 10 min & CNO cycles   \\
\hline $^{16}O\left( p,\gamma \right) ^{17}F$ & 65 s &   \\ \hline
$^{13}N\left( p,\gamma \right) ^{14}O$ & 70.6 s &    \\ \hline
$^{20}Ne\left( p,\gamma \right) ^{21}Na$ & 22.5 s &    \\ \hline
$^{11}C\left( p,\gamma \right) ^{12}N$ & 11 ms & Hot p-p chain  \\
\hline
$^{15}O\left( \alpha ,\gamma \right) ^{19}Ne$ & 17.2 s & rp process   \\
\hline $^{31}S\left( p,\gamma \right) ^{32}Cl$ & 291 ms &    \\
\hline $^{12}C\left( \alpha ,\gamma \right) ^{16}O$ & Stable &
Helium burning  \\ \hline $^{16}O\left( \alpha ,\gamma \right)
^{20}Ne$ & Stable &
\\ \hline
$^{14}N\left( \alpha ,\gamma \right) ^{18}F$ & 109.7 min &    \\
\hline
$^{22}Mg\left( p,\gamma \right) ^{23}Al$ & 3.86 s & rp bottlenecks   \\
\hline
\end{tabular}
\end{center}
\end{table}
\normalsize

The Coulomb dissociation method is more useful when higher order
effects are under control, so that the eq. \ref{Sfactor}, obtained
in 1st-order perturbation theory, is valid. Higher order effects
can be taken into account in a coupled channels approach, or by
using higher order perturbation theory.

Expanding the nuclear wave function in the set $\{\mid j\rangle;\
j=1,N\}$ of eigenstates of the intrinsic Hamiltonian $H_{0}$,
where $N$ is the number states included in the calculation, one
obtains the coupled equations for the occupation amplitude of the
state $k$ as
\begin{equation}
i\hbar\ {\dot{a}}_{k}(t)=\sum_{j=1}^{N}\ \langle k\mid V(t)\mid
j\rangle \;\exp\left[  i(E_{k}-E_{j})t/\hbar\right]
\;a_{j}(t)\;,\qquad\qquad
k=1\;\mathrm{to}\;\;N\ ,\label{AW}%
\end{equation}
where $E_{n}$ is the energy of the state $\left|  n\right\rangle
.$ This set of equations can be solved numerically if one has a
theoretical model or experimental information on the matrix
elements $\langle k\mid V(t)\mid j\rangle$. In the most relevant
situations the perturbing potential is the Coulomb interaction of
the fragments with the target. This interaction is expanded into
electric and magnetic multipolarities. One thus needs information
on the electromagnetic matrix elements between the bound states
and the continuum states of the nucleus to carry out the
coupled-channels calculation. This approach has been used in refs.
\cite{BCa92,CB95} to treat the problem of Coulomb reacceleration
of fragments following the breakup. More recently, a similar
technique has been used in ref. \cite{NT99}.

Another approach to the Coulomb postacceleration, or
reacceleration, problem is to integrate the time-dependent
Schr\"{o}dinger equation directly for a given model Hamiltonian.
This approach is only useful when one can use a two-body potential
model for the nucleus. Then, expanding the two-body wavefunction
into angular components one gets the time-dependent wave equation
\begin{equation}
\left[ {d^2\over dr^2} -{l(l+1) \over r^2} - {2\mu_{bc} \over
\hbar} V(r)\right]u_{lm}(r) + \sum_{l'm'} S_{l'm'}^{lm}u_{l'm'}(r)
= - {2\mu_{bc} \over \hbar} {\partial u_{lm} \over
\partial t}
\end{equation}
where  $u_{lm}(r)$ is the radial part of the $lm$-component of the
two-body wavefunction. $S_{l'm'}^{lm}$ is a source term for the
postacceleration, arising from the multipole component $lm$ of the
Coulomb field of the target. This approach has been used in ref.
\cite{BBe93} and later developed in refs.
\cite{EBB95,KYS96,Mel99,paris99,typwo99}.

As an example of application of the Coulomb dissociation method,
we show in figure \ref{fig4} the result of an experiment performed
at the GSI laboratory, in Darmstadt, Germany \cite{Iw99}. The
S-factor obtained in this experiment is shown in figure \ref{fig4}
as solid circles, by a direct application of eq. \ref{Sfactor},
using $S_{17}(E)$ obtained from theoretical models. The solid
curve is a fit using a theoretical model of ref. \cite{CB95},
whereas the dashed curve is a calculation done in ref.
\cite{DB94}.

Alternative experiments have obtained $S_{17}$ by measuring the
parallel momentum distribution of the fragments after Coulomb
breakup \cite{EB96,Da01}. This technique was applied to several
experiments at the NSCL laboratory, of Michigan State University
\cite{Da01,Dav98,Dav01}. The technique also allows the
disentanglement of the E1 and E2 contributions to the breakup by
looking at the asymmetries in the momentum distribution following
the Coulomb breakup of the projectile \cite{EB96}.

In table I a set of reactions are shown which can be studied with
the Coulomb dissociation method \cite{BR96}.

Besides the reaction $^{7}Be\left( p,\gamma \right) ^{8}B$, the
Coulomb dissociation method has also been applied to the study of
$^{4}He\left( d,\gamma \right) ^{6}Li$ (see ref. \cite{Kie93}),
$^{12}C\left( n,\gamma \right) ^{13}C$ (see ref.
\cite{Kie91,Mot91}), $^{11}C\left( p,\gamma \right) ^{12}N$ (see
ref. \cite{Lef95}), and $^{12}C\left( \alpha ,\gamma \right)
^{16}O$ (see ref. \cite{Tat95}).

\subsection{Multiphonon Resonances}

Giant dipole resonances (GDR) occur in nuclei at energies around
10-20 MeV. Assuming that they are harmonic vibrations of protons
against neutrons, one expects that DGDRs (Double Giant Dipole
Resonances), i.e., two giant dipole vibrations superimposed in one
nucleus, will have exactly twice the energy of the GDR
\cite{BB88,Au98,BP99}.

Assuming that one knows $\sigma^{GDR}_\gamma (E)$ somehow (either
from experiments, or from theory), a simple harmonic model can be
formulated to obtain the Coulomb excitation cross sections of
these states.  In this model the inclusion of the coupling between
all multiphonon states can be performed analytically \cite{BB86}.
Following the same reasoning leading to eq. \ref{epa}, to
first-order the probability to excite a giant resonance state at
energy $E$ is given by
\begin{equation}
P^{1st}(E,b) = {N(E,b) \over E} \ \sigma^{GDR}_\gamma (E) \ ,
\label{PGDR}
\end{equation}
where $N(E,b)$ is the number of equivalent photons per unit area
in a collision at impact parameter $b$. The relation between
$n(E)$ of eq. \ref{epa} and $N(E,b)$ is given by $n(E) = 2\pi \int
N(E,b) \ b\ db$.

In the harmonic oscillator model the excitation probabilities
calculated to first-order, are modified to include the flux of
probability to the other states \cite{BB88}. That is,
\begin{equation}
P(E,\;b)=P^{1st}(E,b)\;\exp \left\{ -P^{1st}(b)\right\} \;,
\end{equation}
where $P^{1st}(b)$ is the integral of $P(E,b)$ over the excitation
energy $E$. In general, the probability to reach a multiphonon
state with the energy $E^{(n)}$ from the ground state, with energy
$E^{(0)},$ is obtained by an integral over all intermediate
energies \cite{LB92}

\begin{eqnarray}
&&P^{(n)}(E^{(n)},b)=\frac 1{n!}\;\exp \left\{
-P^{1st}(b)\right\} \int dE^{(n-1)}\;dE^{(n-2)}\;...\;dE^{(1)}
\\
&&\times P^{1st}(E^{(n)}-E^{(n-1)},b)\;P^{1st}(E^{(n-1)}-E^{(n-2)},b)\;...\;
P^{1st}(E^{(1)}-E^{(0)},b)  \ .
\end{eqnarray}
Integrating this equation over impact parameters yields the cross
section for the excitation of the multiphonon state $n$, as a
function of the excitation energy $E^{(n)}$.

\begin{figure}[t]
\centerline{\psfig{file=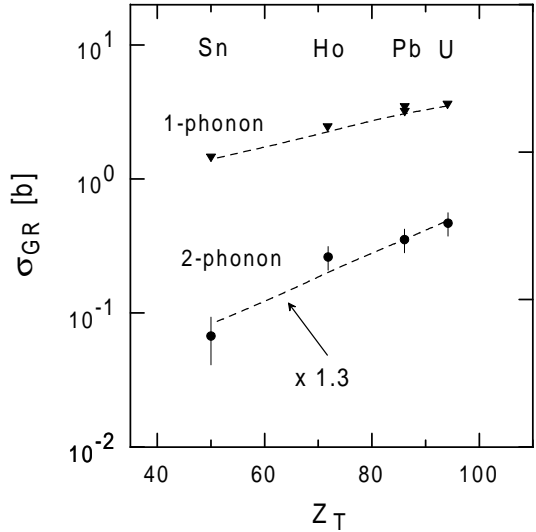,width=7cm}}
\vspace*{8pt}\caption{\small Cross sections for the excitation of
the GDR and the DGDR for several systems. The dashed curves are
theoretical calculations. See text for more details.}
\label{fig5x}
\end{figure}

A series of experiments at the GSI laboratory have obtained the
energy spectra, cross sections, and angular distribution of
fragments following the decay of the DGDR
\cite{Ri93,Sc93,Au93,Bor96,Grun99,Bor99,Ili01}. It was shown that
the experimental cross sections are about 30\% bigger than the
theoretical ones. This is shown in figure \ref{fig5x} where the
cross sections for the excitation of 1-phonon (GDR),
\begin{equation}
\sigma _{1}\sim 2\pi S\ln \left[ {2\gamma A_{T}^{1/3} \over
A_{P}^{1/3}+A_{T}^{1/3}}\right] \ ,
\end{equation}
while for the 2-phonon state
it is
\begin{equation}
\sigma \sim 0.1 \ {S^{2}\over \left(
A_{P}^{1/3}+A_{T}^{1/3}\right) ^{2} \ {\rm mb}}\ ,
\end{equation}
where
\begin{equation}
S=5.45\times 10^{-4}{Z_{P}^{2}Z_{T}N_{T}\over A_{T}^{2/3}} \ \ \
{\rm mb}\ .
\end{equation}
The dashed lines of figure \ref{fig5x} are the result of more
elaborate calculations
\cite{Au98,BP99,Bert96,HTV99,Tol99,Pau01,Psh01}.

The GSI experiments are very promising for the studies of the
nuclear response in very collective states. One should notice that
after many years of study of the GDRs and other collective modes,
the width of these states are still poorly explained
theoretically, even with the best microscopic approaches known
sofar. The extension of these approaches to the study of the width
of the DGDRs will be helpful to improve such models \cite{BP99}.

In heavy ion colliders  the mutual Coulomb excitation of the ions
(leading to their simultaneous fragmentation) is a useful tool for
beam monitoring \cite{seb1}. A recent measurement at RHIC
\cite{seb2}, using the Zero Degree Calorimeter to measure the
neutron decay of the reaction products, has proved the feasibility
of the method. The theoretical prediction of about 3 b for this
process, agrees quite well with the experimental results.

The DGDR contributes to only about 10\% of the total fragmentation
cross section induced by Coulomb excitation in PHIC. The main
contribution arises from the excitation of a single GDR, which
decays mostly by neutron emission. This is also a potential source
of beam loss in relativistic heavy ion colliders \cite{BB94}, and
an important fragmentation mode of relativistic nuclei in cosmic
rays.

\section{Atomic Processes}

\subsection{Atomic ionization}

\begin{figure}[t]
\centerline{\psfig{file=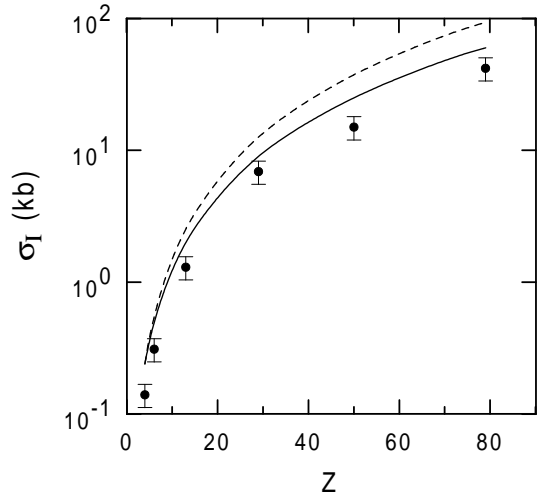,width=7cm}} \vspace*{8pt}
\caption{ Atomic ionization cross sections for $Pb^{81+}$ (33 TeV)
projectiles impinging on several targets. The solid and dashed
curves are theoretical calculations. See text for more details. }
\label{fig2x}
\end{figure}

The cross sections for atomic ionization in PHIC are very large,
of order of kilobarns, increasing slowly with the logarithm of the
RHI energy. For a fixed target experiment using naked projectiles
one gets \cite{BB88} (here we use $\hbar=c=1$)
\begin{equation}
\sigma _{I}=\zeta(3)\ {Z_{P}^{2}r_{e}^{2}\over \left( Z_{T}\alpha
\right)^2} \left[ 1.8\pi +9.8\ln \left( {2\gamma \over
Z_{T}\alpha}\right) \right] \label{atom1}
\end{equation}
which decreases with the target charge $Z_{T}$. This is due to the
increase of the binding energy of $K$-electrons with the atomic
charge. The probability to eject a $K$-electron is much larger
than for other atomic orbitals. In this equation $r_e=e^2/m$ is
the classical electron radius, and $\zeta(3)=1.202$ accounts for
the ionization of electrons from higher orbits.

The first term  inside brackets in eq. \ref{atom1} is due to close
collisions assuming elastic scattering of the electron off the
projectile, while the second part is for distant collisions, with
impact parameter larger than the Bohr radius  \cite{BB88}.
Recently, Baltz \cite{Bal01} has shown that the numerical factors
in the equation above should be replaced by $1.8 \rightarrow
1.74-1.83$ and $9.8 \rightarrow 7.21$, respectively, when one
treats the electronic wavefunctions with the relativistic
corrections.

Recent experiments have reported ionization cross sections for
$Pb^{81+}$ (33 TeV) beams on several targets \cite{Kr98}. In this
case, the role of projectile and target are exchanged in the
previous equation. In figure \ref{fig2x} show the results of  eq.
\ref{atom1} (dashed line) are compared to the experimental data.
Since the targets are screened by their electrons, the discrepancy
is expected. Calculations by Anholt, Becker and collaborators
\cite {AB87,EM95,AG86} (solid line) or of Baltz \cite{Bal01} also
yield larger cross sections than the experimental data.

Non-perturbative calculations, solving the time dependent Dirac
equation exactly, were first performed by Giessen and Oak Ridge
groups \cite {Bec83,BS85}. The main problem is to adequately treat
the several channels competing with the ionization process,
specially for atoms with more than one electron. Also, the effects
of screening (static and dynamic) are hard to calculate. On the
experimental side, there are little data available for a
meaningful comparison with theory. However, atomic ionization
should be taken seriously in accelerator design when a residual
gas remains in the accelerator pipes.

\subsection{Bremsstrahlung and Delbr\"{u}ck scattering}

Bremsstrahlung (fig. \ref{graphs5}c) is a minor effect in PHIC. It
has been calculated in ref.  \cite{BB88,BB89}. The cross section
is proportional to the inverse of the square mass of the ions.
Most virtual photons have very low energies. For 10 MeV photons
the central collisions deliver 10$^{6}$ more photons than the
peripheral ones \cite{BB89}.

For a
collider the Bremsstrahlung differential cross section is given by
\begin{equation}
{d\sigma
_{\gamma }\over d\omega} ={16Z^{6}\alpha ^{3}\over  3\omega A^{2}m_{N}^{2}}
\ln \left( {\gamma \over \omega R}\right)
\ , \label{delb}
\end{equation}
where $m_{N} $ is the nucleon
mass, $\gamma =2\gamma _{c}^{2}-1$, where $\gamma _{c} $ is the collider
Lorentz gamma factor ($\gamma _{c}\sim 100$ for RHIC/BNL), and $R$ is the
nuclear dimension ($R\sim 2.4\times A^{1/3}$ fm) \cite{BB89}.

For very low energy photons ($\omega \sim 100$ eV) the whole set
of particles in a beam bunch can act coherently and a great number
of Bremsstrahlung photons can be produced. This has been proposed
as a tool for monitoring the bunch dimension in colliders
\cite{Gin92}. Recently, an experiment has been approved at
RHIC/BNL to measure this kind of coherent effect, including the
emission of real photons by the interaction of a bunch with the
edges of the accelerator magnets \cite{serbo02}.

Delbruck scattering ($\gamma ^{\ast }+\gamma ^{\ast
}\longrightarrow \gamma +\gamma $) involves an additional $\alpha
^{2}$ as compared to pair production and has never been possible
to study experimentally. \ The cross section is about 50 b for the
LHC \cite{BB89} and the process is dominated by high-energy
photons, $E_{\gamma }\gg m_{e}$. A study of this process in PHIC
is thus promising if the background problems arising from central
collisions can be eliminated. No experiments of Bremsstrahlung or
Delbr\"{u}ck scattering in PHIC have been performed so far. The
total cross section for Delbr\"{u}ck scattering ($\omega \gg
m_{e}$) in colliders is given by  \cite{BB89}
\begin{equation}
\sigma =2.54Z^{4}\alpha ^{4}r_{e}^{2}\ln ^{3}\left( {\gamma \over
m_{e}R}\right) \ . \label{delb2}
\end{equation}

\section{The Electromagnetic Field of an Ultrarelativistic
Charge}

As discussed above, the calculation of processes involving the
multiple exchange of photons is not as easy to perform. For
example,  substantial differences  have been found between the
cross sections for the production of $e^{+}e^{-}$  pairs
calculated  within several approaches \cite{ERG00,BGMP01}.
However, the calculation of multiple photon exchange can be
considerably simplified at very high bombarding energies if one
assumes that the electromagnetic field generated by the projectile
(or target, depending on the reference frame) lies entirely on a
plane perpendicular to its trajectory \cite{Ba91}. We show next
that this approximation is well justified \cite{Ber01}.

The scalar part of the electromagnetic potential generated by an
ultrarelativistic particle is proportional to $V\left(
\mbox{\boldmath$\rho$} ,z,t\right) =\phi \left( \mbox{
\boldmath$\rho$} ,z,t\right)$ where $\phi \left(
\mbox{\boldmath$\rho$} ,z,t\right) $ is the Lienard-Wiechert
potential at a point ${\bf r}=\left( \mbox{\boldmath$\rho$}
,z\right) $, generated by a relativistic particle with velocity
${\bf v}=v\hat{\bf z}$ and impact parameter ${\bf b}$ (in our
units $\hbar =c=m_{e}=1 $),

\begin{equation}
\phi \left( \mbox{\boldmath$\rho$} ,z,t\right) =\gamma Z\alpha \left[ \left( {\bf b}-\mbox{\boldmath$\rho$} \right)
^{2}+\gamma ^{2}\left( z-vt\right) ^{2}\right] ^{-1/2}\;.\label{phi}
\end{equation}
In these expressions $\gamma =\left( 1-v^{2}\right) ^{-1/2}$ is
the Lorentz contraction factor. The vector part of the
electromagnetic interaction is proportional to eq. \ref{phi} and ,
for simplicity, we leave it out of this proof.

Using the Bethe-integral \cite{BB88}, the potential (\ref{phi}) can be written in the form

\begin{equation}
\phi \left( \mbox{\boldmath$\rho$} ,z,t\right) =Z\alpha \, \frac{1}{2\pi ^{2}}\int d^{3}q\frac{%
e^{-i{\bf q.u}}e^{i{\bf q.r}}}{q^{2}-v^{2}q^{2}_{z}}\, , \label{bethe}
\end{equation}
where ${\bf u}={\bf b}+{\bf v}t$ and $q=\left( {\bf q}_{t},\  q_{z}\right) $.
The integral in (\ref{bethe}) diverges logarithmically as $v\,
\rightarrow \, 1 $.

For practical purposes one needs the interaction in the form
$V\left( \mbox{\boldmath$\rho$} ,z,t\right)=\phi\left(
\mbox{\boldmath$\rho$} ,z,t\right)-\phi\left( 0 ,z,t\right)$,
i.e.,
\begin{equation}
V\left( \mbox{\boldmath$\rho$} ,z,t\right) =Z\alpha \,
\, \frac{1}{2\pi ^{2}}\int d^{3}q\ e^{-i{\bf {\bf q.u}}}\frac{\left[ e^{i{\bf q.r}}-e^{iq_{z}z}%
\right] }{q_{t}^{2}+q^{2}_{z}/\gamma ^{2}}\, . \label{full}
\end{equation}

Let us define
\begin{eqnarray}
\Phi \left( \mbox{\boldmath$\rho$} ,z,t\right) &\equiv& {V\left( \mbox{\boldmath$\rho$} ,z,t\right)
\over Z\alpha \, } \nonumber \\
&=&\frac{1}{\pi }\int d^{2}q_{t}\, \frac{1}{
q^{2}_{t}}\, \exp{(-i{\bf q}_{t}.{\bf b})}\, \left[ \exp{(i{\bf q}_{t}.\mbox{\boldmath$\rho$}) }-1
\right] \, \Phi
_{z}\left( {\bf q}_{t},z,t\right) \, . \label{rhoz}
\end{eqnarray}
Then
\begin{equation}
\Phi _{z}\left( {\bf q}_{t},z,t\right) =\frac{q^{2}_{t}}{2\pi }\int dq_{z}\frac{%
e^{iq_{z}\left( z-vt\right) }}{q_{t}^{2}+q^{2}_{z}/\gamma ^{2}}=\frac{\gamma
q_{t}}{2}e^{-\gamma q_{t}|z-vt|}\, .
\end{equation}
Now, using $\lim_{\Lambda \rightarrow \infty }\, \left(
\Lambda /2\right) e^{-\Lambda |x|}=\delta \left( x\right) $, we see that for
$\gamma \rightarrow \infty $, $\Phi _{z} $ does not depend
on $q_{t} $, and assumes the form of a delta function: $\Phi _{z}\left(
z,t\right) =\delta \left( z-vt\right) $.

In this limit, we can write (\ref{rhoz})  as
\begin{equation}
\Phi \left( \mbox{\boldmath$\rho$} ,z,t\right) =\delta \left( z-t\right)
\Phi_\rho \left( \mbox{\boldmath$\rho$}
,t\right) \, ,
\end{equation}
with
\begin{eqnarray}
\Phi_\rho \left( \mbox{\boldmath$\rho$} ,t\right) &=&\frac{1}{\pi }\int d^{2}q_{t}\,
\frac{1}{q^{2}_{t}}\, \exp{(-i{\bf q}_{t}.{\bf b})}\, \left[ \exp{(i{\bf q}_{t}.
\mbox{\boldmath$\rho$}) }-1\right]
\nonumber \\
&=&2\int \frac{dq_{t}}{q_{t}}\left\{ J_{0}\left[ q_{t}|
\mbox{\boldmath$\rho$} -b|\right] -J_{0}\left(
q_{t}b\right) \right\} \, ,
\end{eqnarray}
where $J_{0} $ is the cylindrical Bessel function. The integral
over each Bessel function diverges, but their difference does not.
To show this one regularizes the integrals by using

\begin{equation}
\int dx\, \frac{xJ_{0}\left( ax\right) }{x^{2}+k^{2}}=K_0\left( ak\right) \, ,
\end{equation}
where $K_{0} $ is the modified cylindrical Bessel function. Taking
the limit $k\rightarrow 0 $, and using $K\left( ak\right) \simeq
\ln\left( ak\right) $, for small values of $ak$, one gets

\begin{equation}
\Phi \left( \mbox{\boldmath$\rho$} ,t\right) =
\ln\frac{\left( {\bf
b}-\mbox{\boldmath$\rho$} \right) ^{2}}{b^{2}}\, .
\end{equation}

This is the solution of the Coulomb potential of a unit charge in
2-dimensions. An easy way to see this is to use Gauss law for the
electric field in two dimensions. One obtains $E\simeq 1/b $,
where $b $ is the distance to the charge. Since $E=-\partial \Phi
/\partial b $, the logarithmic form of $\Phi $ is evident.

The above derivation illustrates the validity of the approximation
in terms of the transverse momentum transfer $q_{t} $. It should
fail for very soft processes, i.e., those for which $q_{t}
$$\rightarrow 0. $ Also, it requires that $q_{z} $ is small
compared to $\gamma $. As shown in ref. \cite{BB88}, $q_{z} $
values  up to $\gamma$ units of the electron mass contribute
appreciably to the integrals involved in the production of free,
and of bound-free,  $e^{+}e^{-} $ pairs.

The delta-function interaction yields reasonable results as long
as $\omega b/\gamma \leq 0.1.$ \cite{Ber01}. This amounts to
$b\leq 0.1\gamma /\omega $. As shown in ref. \cite{BB88}, the most
effective impact parameters for this process are of the order of \
$b \simeq 1/m$. In ref. \cite{Ber01} it has been shown that the
differential cross sections for $e^{+}e^{-} $ pair production are
well described up to energies of the order of $0.1\gamma m$, by
using the delta-function interaction.

For other situations, e.g., nuclear fragmentation due to the
electromagnetic interaction, the most effective impact parameter
is given by $b \simeq R$, where $R\simeq 10$ fm. We thus expect
that the delta-function interaction works well for $\varepsilon
\leq 0.1\gamma $ MeV. Note that  $\gamma$ is huge for RHIC and LHC
energies, and the approximation works well for all energies of
practical interest in nuclear fragmentation.

The basic idea of the delta-function interaction is that the
electromagnetic field of a relativistic charge looks like a very
thin pancake. Those processes which do not involve too large
energy transfers, will not be sensitive to the spatial variation
of the field. Then the delta-function is a good approximation. For
typical energy transfers of the order of 10-100 MeV in nuclear
fragmentation, the approximation works well for $b\leq 0.1\gamma
/\omega $ fm. To calculate total cross sections it is always
necessary to account for those large impact parameters at which
the delta-function approximation fails. Similar conclusions have
been drawn in a recent article on projectile-electron loss in
relativistic collisions with atomic targets \cite{Vo00}.

\subsection{Free and bound-free electron-positron pair production}

It has been demonstrated long time ago \cite{Fu33,LL34,Bha35,NTK35} that to leading order in $%
\ln \gamma $, the $e^+e^-$-pair production in PHIC is given by
\begin{equation}
\sigma
={28\over 27\pi} \ Z_{P}^{2}Z_{T}^{2}r_{e}^{2}\ln ^{3}\left( {
\gamma \over 4}\right)
\ .
\end{equation}

A renewed interest in this process appeared with the construction
of relativistic heavy ion accelerators. For heavy ions with very
large charge (e.g, lead, or uranium) the pair production
probabilities and cross sections are very large. They cannot be
treated to first order in perturbation theory \cite{BB88}, and are
also difficult to calculate. This resulted in a great number of
theoretical studies \cite{RB91,Ba90,RBW91,BGS92,Vid93,Hen98}.

\begin{figure}[t]
\centerline{\psfig{file=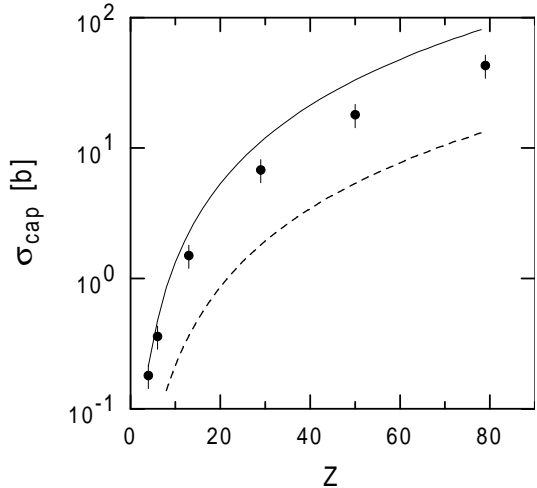,width=7cm}} \vspace*{8pt}
\caption{ Pair production with capture for $Pb^{82+}$ (33 TeV)
beams on several targets. The solid and dashed curves are
theoretical calculations. See text for more details.} \label{fig3}
\end{figure}

Replacing the Lorentz compressed electromagnetic fields by delta functions,
and working with light cone variables, one has developed more elaborate
calculations \cite{Ba97}, recently. One still debates upon
the proper treatment of Coulomb distortion of the lepton wavefunctions, and
of production of n-pairs \cite{Ba97}.

An important phenomenon occurs when the electron is captured in an
atomic orbit of the projectile, or of the target. In a collider
this leads to beam losses each time a charge modified nucleus
passes by a magnet downstream \cite{BB89}. A striking application
of this process was the recent production of antihydrogen atoms
using relativistic antiproton beams \cite {Ba96,Bla80}. Here the
positron is produced and captured in an orbit of the antiproton.
Early calculations for this process used perturbation theory
\cite{BB88,AB87,Bec87,Mom91,Rum93,Mom95}. Some authors use
non-perturbative approaches, e.g., coupled-channels calculations
\cite{RB91,Ba90,RBW91,BGS92,Vid93,Hen98}. Initially some
discrepancy with perturbative calculations were found, but later
it was shown that non-perturbative calculations agree with the
perturbative ones at the 1\% level (see, e.g., ref.
\cite{Ba97,Bal98,SW98,Iva99,Eic99,BGMP01,LeM01}).

The expression
\begin{equation}
\sigma ={3.3\pi Z^{8}\alpha ^{6}r_{e}^{2}\over \exp \left(
2\pi Z\alpha \right) -1} \left[ \ln \left( {0.681 \gamma\over  2}
\right) -{5\over 3}\right]
\label{BF}
\end{equation}
for pair production with electron capture in PHIC was obtained in
ref. \cite{BB88}. The term in the denominator is the main effect
of the distortion of the positron wavefunction. It arises through
the normalization of the continuum wavefunctions which accounts
for the reduction of the magnitude of the positron wavefunction
near the nucleus where the electron is localized (bound). Thus,
the greater the $Z$, the less these wavefunctions overlap. Eq.
\ref{BF} predicts a dependence of the cross section in the form
$\sigma =A\ln \gamma +B$, where $A$ and $B$ are coefficients
depending on the system. This dependence was used in the analysis
of the experiment in ref. \cite{Kr98}.

In recent calculations, attention was given to the correct
treatment of the distortion effects in the positron wavefunction
\cite{He00}. In figure \ref{fig3} we show the recent experimental
data of ref. \cite{Kr98} compared to eq. \ref{BF} and recent
calculations \cite{BD00,He00}. These calculations also predict a
$\ln \gamma$ dependence but give larger cross sections than in
ref. \cite{BB88}. The comparison with the experimental data is not
fair since atomic screening was not taken into account. When
screening is present the cross sections will always be smaller up
to a factor of 2 (see ref.  \cite{BB88}). The conclusion here is
that pair production with electron capture is a process which is
well treated in first order perturbation theory. However, eq.
\ref{BF} is shown to be only a rough estimate of the cross section
\cite{BD00}. As with the production of free pairs, the main
concern here is the correct treatment of distortion effects
(multiphoton scattering) \cite{BD00,He00}.

The production of para-positronium in heavy ion colliders was
calculated \cite{Kot99}. The cross section at RHIC is about 18 mb.
This process is of interest due to the unusual large transparency
of the parapositronium in thin metal layers.

The production of mesons in two- and three-photon collisions can
be studied following the same formalism as used in QED for the
production of positronium \cite{BN02}. This is shown in the next
section.

\section{Photon-photon Fusion Mechanism}

\subsection{Meson Production}

The production of heavy lepton pairs ($\mu^+\mu^-$, or $\tau^+\tau^-$), or
of meson pairs (e.g., $\pi^+\pi^-$) can be calculated using the second of
equation (1). One just needs the cross sections for $\gamma\gamma$
production of these pairs. Since they depend on the inverse of the square of
the particle mass \cite{BB89}, the pair-production cross sections are much
smaller in this case. The same applies to single meson production by $
\gamma\gamma$ fusion.
Low \cite{Low60}
showed that one can relate the particle production
by two real photons (with
energies $\omega_1$ and $\omega_2$, respectively)
to the particle's decay width, $\Gamma_{\gamma\gamma}$. Since both
processes involve the
same matrix elements, only the
phase-space factors and polarization summations are
distinct. Low's formula is
\begin{equation}
\sigma (\omega_1, \ \omega_2 ) = 8\pi^2 (2J+1){\Gamma_{\gamma\gamma} \over M} \
\delta (4\omega_1 \omega_2 - M^2) \ ,
\label{photon}
\end{equation}
where $J$, $M$,
and $\Gamma_{M\rightarrow\gamma\gamma}$ are the spin, mass and two-photon
decay width of the meson, $W$ is the c.m. energy of the colliding photons
\cite{BB89}.
The delta-function accounts for energy conservation.

In ref. \cite{BB88} the following equation was obtained for the
production of mesons with mass $M$ in HI colliders:
\begin{equation}
\sigma = {128\over 3} \ Z^4\alpha^2 {\Gamma_{\gamma\gamma}\over
M^3} \ \ln^3 \left( {2\gamma \delta \over M R} \right) \ ,
\label{ln}
\end{equation}
where $\delta=0.681...$ Later \cite{BF90} it was shown that a more
detailed account of the space geometry of the two-photon collision
is necessary, specially for the heavier mesons. $R$ is a parameter
which depends on the mass of the produced particle. If $M$ is much
smaller than the inverse of a typical nuclear radius,  then $R =
1/M$, otherwise $R$ is the nuclear radius. These choices reflect
the uncertainty relation in the direction transverse to the beam,
as explained in ref. \cite{BB88}. Since spin 1 particles cannot
couple to two real photons \cite{Ya50}, one expects that only spin
0 and spin 2 particles are produced.

In the next section we show how one can calculate meson production
using the concepts of QED extended to include bound quark states
\cite{BN02}.

\begin{figure}[t]
\centerline{\psfig{file=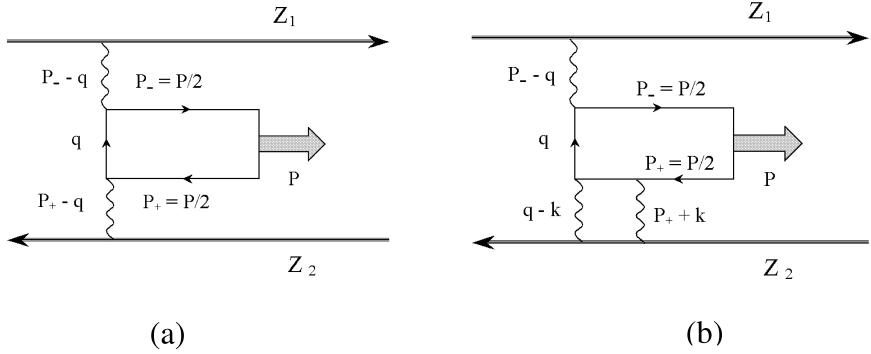,width=13cm}}
\vspace*{8pt}\caption{\small Feynman graphs for two- and
three-photon fusion in ultra-peripheral collisions of relativistic
heavy ions. } \label{f1}
\end{figure}

\subsection{Two-photon fusion in heavy ion colliders}

In the laboratory frame the Fourier components of the classical electromagnetic field
at a distance ${\bf b}/2$ of
nucleus 1 with charge $Ze$ and velocity $\beta$, is given by (in our notation
$q = (q_0,{\bf q}_t, q_3)$, and $q_3 \equiv q_z$)
\begin{equation}
A^{(1)}_0(q)=-8\pi^2Ze\ \delta(q_0-\beta q_3){e^{i{\bf q}_t .{\bf b}/2}
\over {q^2_t +q_3^2/\gamma^2}}\ \ \ \ \ \ \
{\rm and} \ \ \ \ A^{(1)}_3 = \beta A^{(1)}_0
\label{EMF}
\end{equation}
For the field of nucleus 2, moving in the opposite direction, we replace $\beta$
by $-\beta$   and ${\bf b}$ by $-{\bf b}$ in the equations above. Although
$\beta \simeq 1$ in relativistic colliders, it is important to keep
them in the key places,
as some of their combinations will lead to important
$\gamma = (1-\beta^2)^{-1/2}$ factors.

The matrix element for the production of positronium is
directly obtained from the
corresponding matrix element
for the production of a free pair (see fig. \ref{f1}(a)), with the requirement that
$P_+=P_-=P/2$, where $P$ is the momentum of the final bound state.
That is
\begin{eqnarray}
{\cal M} &=&
{\cal M}_1 +{\cal M}_2 \nonumber \\
&=& -i e^2 \bar{u}({P\over 2})\left[ \int{d^4q\over{(2\pi)^4}}
\not\!\!
A^{(1)}({P\over 2}-q)
{{\not\! q + M/2}\over {q^2 - M^2/4}}
\not\!\! A^{(2)}({P\over 2}+q) \right.
\nonumber \\
&+& \left.
\not\!\! A^{(1)}({P\over 2}+q)
{{\not\! q + M/2}\over {q^2 - M^2/4}}
\not\!\! A^{(2)}({P\over 2}-q)
\right] v({P\over 2}) \ ,\label{M1}
\end{eqnarray}
where $M$ is the positronium mass.

The treatment of bound states in quantum field theory is a very
complex subject (for reviews, see \cite{BYG85,Sap90}). In our
case, we want to use the matrix element for free-pair production
and relate the results for the production of a bound-pair. A
common trick used in this situation is to convolute the matrix
element given above with the bound-state wave function. One can
show (see, e.g., \cite{Nov78}) that this is equivalent to the use
of projection operators of the form
\begin{eqnarray}
\bar{u} &\cdots&  v \longrightarrow
\ {\Psi(0)\over 2\sqrt{M}}
\ {\rm tr} \left[\cdots (\not\!\! P + M) i \gamma^5\right] \  ,
\ \ {\rm and} \nonumber \\
\bar{u} &\cdots&  v \longrightarrow
\ {\Psi(0)\over 2\sqrt{M}}
\ {\rm tr} \left[\cdots (\not\!\! P + M) i \not\!\hat{e}^* \right]
\label{traces}
\end{eqnarray}
where $\cdots$ is any matrix operator. The first equation applies
to a spin 0 (parapositronium) and the second to spin 1
(orthopositronium) particles,
respectively. In these equations $\Psi({\bf r})$ is
the bound state wavefunction calculated
at the origin, and $\hat{e}^*$ is the
polarization vector, given by
$\hat{e}^*_{\pm 1}= (0, 1/\sqrt{2}, \pm i/\sqrt{2}, 0)$ and
$\hat{e}^*_0 = (0,0,0,1)$.

Using eq. \ref{traces} in eq. \ref{M1}, one gets for the parapositronium production
\begin{equation}
{\cal M}=16 i {\Psi(0)\over \sqrt{M}} \left( Z\alpha\right)^2 \Big|{\bf P \times I}\Big| \ ,
\end{equation}
where
\begin{equation}
{\bf I} = \int {d^2q_t \ {\bf q}_t \over q_t^2+{\cal Q}^2} \
{1
\over
\left[\left({\bf P}_t/2+{\bf q}_t\right)^2 +\omega_1^2/\gamma^2
\right]}\
{1
\over
\left[\left({\bf P}_t/2-{\bf q}_t\right)^2 +\omega_2^2/\gamma^2
\right]}
\ ,
\end{equation}
with
\begin{equation}
{\cal Q}^2 = {M^2\over 2}+{P_t^2 \over 4} +{P_z^2\over 2 \gamma^2}
\simeq {M^2\over 2}+{P_t^2 \over 4}
\label{calQ}
\end{equation}
and
\begin{equation}
\omega_1 = {E/\beta - P_z\over 2} \ ,\ \  \
\omega_2 = {E/\beta + P_z\over 2} \ \ \ {\rm and}
\ \ \ \ 4\omega_1\omega_2 = M^2+P_t^2-P_z/\gamma^2 \simeq M^2+P_t^2
\ ,
\label{omega12}
\end{equation}
where $E\equiv P_0$ is the total positronium energy.

We see that $\omega_1$ and $\omega_2$ play the role of the (real) photon
energies. For real photons one expects $4\omega_1\omega_2 = E^2$, as
in eq. \ref{photon}.

The two-photon fusion cross sections can be obtained  by using

\begin{equation}
d\sigma = \sum_{\mu} \left[
\int d^2b \ |{\cal M}(\mu)|^2\right]\  {d^3P\over (2\pi)^3 \ 2 E}
\label{sig}
\end{equation}
The positronium wavefunction at the origin is very well known. It
is given by $ \Big|\Psi(0)\Big|^2 = M^3 \alpha^3/64\pi$, where $M$
is the positronium mass.

Since the important impact parameters for the production of the
positronium will be $b > 1/m_e \gg R$, where $R$ is the nuclear
radius, the integral over impact parameter can start from $b=0$.
Thus, the integral over impact parameter in eq. \ref{sig} yields a
delta function. Changing to the variables
\begin{equation} {\bf q}_{1t} = {{\bf P}_t\over 2}- {\bf q}_t \ ,
\ \ \ {\bf q}_{2t} = {{\bf P}_t\over 2}+ {\bf q}_t \ , \ \ \ \
q_{1z}= {P_z/2 - q_z \over \gamma} \  , \ \ \ \ q_{2z}= {P_z/2 +
q_z \over \gamma } \ , \label{qs} \end{equation} eq. \ref{sig}
becomes \begin{equation} E {d\sigma \over d^3 P} ={\zeta (3) \over
\pi} \ {\sigma_0 \over M^2} \ J_B \ , \label{sigJB}
\end{equation}
where
\begin{equation}
J_B = {M^2 \over \pi} \int {\bf A}^2 \delta \left( {\bf q}_{1t}+
{\bf q}_{2t}-{\bf P}_t\right) d{\bf q}_{1t} d{\bf q}_{2t} \ ,
\label{JB}
\end{equation}
with
\begin{equation}
{\bf A} = {{\bf q}_{1t}\times {\bf q}_{2t} \over q_1^2q_2^2} \
{M^2 \over M^2 -q_1^2 -q_2^2} \ ,\ \ \ \ \ \zeta(3) = 1.202... \ ,
\ \ \ \ {\rm and} \ \ \ \ \ \sigma_0 = {4Z^4\alpha^7\over M^2} \ .
\end{equation}
We have included the zeta-function $\zeta(3)$ to take into account
the production of the para-positronium in higher orbits, besides
the production in the K-shell.

The equation \ref{sigJB} is equal to the equation obtained in ref.
\cite{Kot99} for the production of the parapositronium, using an
alternative method.

The integral in eq. \ref{JB} can be carried out analytically,
reducing it to a 3-dimensional integral \cite{BN02}. For RHIC,
using $\gamma = 108$ and $Au+Au$ collisions, one finds $\sigma =
19.4$ mb. For the LHC , using $\gamma = 3000$ and $Pb+Pb$
collisions, one finds $\sigma = 116$ mb. These are in good
agreement with the results (Born cross sections) of ref.
\cite{Kot99}.  Notice however, that Coulomb corrections are very
important, due to the low mass of the electron and the positron
\cite{Kot99}. When Coulomb corrections are included to the Born
cross sections, the final values decrease by as much as 43\% for
RHIC and 27\% for LHC. For meson production one expects that these
corrections ere less relevant.

\subsection{ Production of $C=even$ Mesons}

One can extend the calculation of the previous section to account
for the production of mesons with spin $J = 0$ and $J =2$ by the
two-photon fusion mechanism. The following procedure is to be
adopted:
\begin{enumerate}
\item Replace the electron-positron lines by quark-antiquarks in
the diagram of figure \ref{f1}(a). \item $M$ in the following
formulas will refer to the meson mass. \item Replace $\alpha^2$ by
$\alpha^2 \ (2J+1) \ 3 \sum_i Q_i^4$, where 3 accounts for the
number of colors, and $Q_i$ is the fractional quark charge. These
two last factors will cancel out when one expresses
$\Big|\Psi(0)\Big|^2$ in terms of $\Gamma_{\gamma\gamma}$, the
decay-width of the meson. To understand how this is done, lets
discuss the basics of the annihilation process of a positronium
(see also ref. \cite{BLP82}). With probability $\alpha^2$ the
$e^-$ can fluctuate and emit a virtual photon with energy $~ m_e$.
The electron recoils and can travel up to a distance $\sim 1/m_e$
(or time $\sim m_e$) to meet the positron and annihilate. This
occurs when $e^-$ and $e^+$ are both found close together in a
volume of size $(1/m_e)^3$, i.e., with a probability given by
$|\Psi(0)|^2/m_e^3$. Thus, the annihilation probability per unit
time (decay width) is $\Gamma \sim \alpha^2 |\Psi(0)|^2/m_e^2$.
Angular momentum conservation and CP invariance does not allow the
ortho-positronium to decay into an even number of photons
\cite{Ya50}. The description of the annihilation process given
above is thus only appropriate for the para-positronium. A
detailed QED calculation yields an extra  $4\pi$ in the formula
above. This yields $\Gamma_{\gamma\gamma} (^1S_0)= 8.03 \times
10^9 \ s^{-1}$, while the experimental value \cite{Th70} is
$7.99(11) \times 10^9 \ s^{-1}$, in good agreement with the
theory. For mesons, including the color and the charge factors, as
described before, the relationship between $\Psi(0)$ and
$\Gamma_{\gamma\gamma}$ arise due to the same reasons. One gets
$\Gamma_{\gamma\gamma}=16\pi\alpha^2 \Big|\Psi(0)\Big|^2/M^2 \cdot
3 \sum_i Q_i^4$.

According to these arguments the connection between
$\Gamma_{\gamma\gamma}$ and $\Big|\Psi(0)\Big|^2$, extended to
meson decays, should be valid for large quark masses so that
$1/m_q \ll \ \sqrt{<r^2>}$, where $\sqrt{<r^2>}$ is the mean size
of the meson. Thus, it should work well for, e.g. charmonium
states, $c\bar c$. In fact, Appelquist and Politzer \cite{AP75}
have generalized this derivation for the hadronic decay of heavy
quark states, which besides other phase-space considerations
amounts in changing $\alpha$ to $\alpha_s$, the strong coupling
constant. This can be simply viewed as a way to get a constraint
on the wavefunction $|\Psi(0)|^2$ (see, ref. \cite{RW67}). One
expects that these arguments are valid to zeroth order in quantum
chromodynamics and in addition one should include relativistic
corrections. But, as shown in \cite{BLP82}, the inclusion of
relativistic effects, summing diagrams to higher order in the
perturbation series, is equivalent to solving the non-relativistic
Schr\"odinger equation.

\item Change the integration variable to ${\bf q}_{1t}$ and
${\bf q}_{2t}$.

\item Introduce form factors $F(q_{1t})$ and $F(q_{2t})$ to
account for the nuclear dimensions. This is a simple way to
eliminate the integral over impact parameters and is justified 'a
posteriori', i.e., when one compares the results with those from
other methods. These form factors will impose a cutoff in $q_{1t}$
and $q_{2t}$, so that
\begin{equation}
q_{1t}\ , \ \ q_{2t} \simeq {1\over R} \ll M
\label{llM}
\ ,
\end{equation}
where $R$ is a typical nuclear size. Taking $R = 6.5$ fm, one gets
$1/R \sim 30$ MeV. This is much smaller than the meson masses. As
an outcome of this condition, one can replace ${\cal Q}^2 \sim M^2
/2$ in eq. \ref{calQ}.

\end{enumerate}

According the procedures 1-5 one gets from eq. \ref{sigJB},
\begin{eqnarray}
{d\sigma \over d P_z} &=& {16 (2J+1)\over \pi^2} {Z^4 \alpha^2 \over M^3}
\ \Gamma_{\gamma\gamma} \ {1\over E}\nonumber \\
&\times&
\int d{\bf q}_{1t} d{\bf q}_{2t} \
({\bf q}_{1t}\times {\bf q}_{2t})^2\
{\left[ F_1(q_{1t}^2)F_2(q_{2t}^2)\right]^2 \over
\left(q_{1t}^2+\omega_1^2/\gamma^2\right)^2
\left(q_{2t}^2+\omega_2^2/\gamma^2\right)^2}
\label{meson2}
\end{eqnarray}

Using eqs. \ref{omega12} one has
\begin{equation}
E = \omega_1 + \omega_2 \ , \ \ \ \
\omega_1 - \omega_2 = P_z \ , \ \ \ \ \
{\rm and} \ \ \ \ \
\omega_1 \omega_2 = M^2/4
\nonumber
\end{equation}
so that
\begin{equation}
dP_z = \left( 1 + {M^2 \over 4 \omega_1^2} \right) d\omega_1
\ , \ \ \ \ \ {\rm and} \ \ \ \ \
E =  {\omega_1^2 +M^2/4 \over \omega_1}
\ .
\end{equation}
Thus,
\begin{equation}
{d\sigma \over d \omega_1}  =  \sigma^{(+)} \ {d{\cal N}_{2\gamma}(\omega_1)
\over d\omega_1}
= \sigma^{(+)}
 \ {1\over \omega_1}
\
n_1(\omega_1) n_2(\omega_2) \ ,
\label{epaf1}
\end{equation}
where
\begin{equation}
\sigma^{(+)} = 8 \pi^2 (2J+1) {\Gamma_{\gamma\gamma} \over M^3}
\ \ \ \ {\rm and} \ \ \ \ \
n_i(\omega_i) =  {2\over \pi}\ Z^2 \alpha \ \int {dq \ q^3
\ \left[ F_i(q^2) \right]^2 \over
\left(q^2+\omega_i^2/\gamma^2\right)^2} \ .
\label{epaf2}
\end{equation}

We notice that $n(\omega)$ is the frequently used form of the
equivalent photon number which enters eq. \ref{epa}. Thus, eqs.
\ref{epaf1} and \ref{epaf2} are the result one expects by using
the equivalent photon method, i.e., by using eqs. \ref{photon} and
\ref{epa}. This is an important result, since it shows that the
projection method to calculate the two-photon production of mesons
works even for light quark masses (i.e., for $\pi^0$). In this
case there seems to be no justification for replacing the quark
masses and momenta by half the meson masses and momenta, as we did
for the derivation of eq. \ref{meson2}. This looks quite
intriguing, but it is easy to see that the step 3 in our list of
procedures adopted is solely dependent on the meson mass, not on
the quark masses, i.e., if they are constituent, sea quarks, etc.
Moreover, the projection method eliminates the reference to quark
masses in the momentum integrals. The condition \ref{llM} finishes
the job, by eliminating the photon virtualities and yielding the
same result one would get with the equivalent photon
approximation.

One can define a ``two-photon equivalent number'', ${\cal
N}_{2\gamma}(M^2)$, so that $\sigma = \sigma^{(+)} {\cal
N}_{2\gamma}(M^2)$, where
\begin{equation}
 {\cal N}_{2\gamma}(M^2) =\int d\omega {d{\cal N}_{2\gamma} \over d \omega}=
\int {d\omega\over \omega} \ n_1(\omega)
\ n_2\left({M^2\over 4\omega}\right) \ .
\end{equation}

To calculate this integral one needs the equivalent photon numbers
given by eq. \ref{epaf2}. The simplest form factor one can use for
this purpose is the `sharp-cutoff' model,  which assumes that
\begin{equation}
F(q^2) = 1 \ \ {\rm for} \ \
q^2 < 1/R \ ,
\ \ \ {\rm and }\ \ \ F(q^2) = \ 0 \ ,
\ \ {\rm otherwise} \ .
\end{equation}

In this case, one gets for the differential cross section
\begin{eqnarray}
{d\sigma \over d \omega} &=& \sigma^{(+)} \ {Z^4\alpha^2\over \pi^2\omega}
\ \left[
\ln \left( 1+{\gamma^2 \over \omega^2 R^2} \right)-
{1\over 1+\omega^2 R^2/\gamma^2}\right]
\nonumber \\
&\times&\left[
\ln \left( 1+{16\gamma^2  \omega^2 \over M^4R^2} \right)-
{1\over 1+  M^4R^2/ 16\gamma^2  \omega^2  }\right]
\label{LLspec2}
\ .
\end{eqnarray}
The spectrum possesses a characteristic $1/\omega$ dependence, except
for $\omega \gg \gamma/R$, when it decreases
as $1/\omega^5$.

When the condition $\gamma \gg M R$ is met, one can neglect the
unity factors inside the logarithm in eq. \ref{LLspec2}, as well
as the second term inside brackets. Then, doing the integration of
\ref{LLspec2} from $\omega = M^2R/4/\gamma$ to $\omega =
\gamma/R$, we get eq. \ref{ln}. But eq. \ref{LLspec2} is an
improvement over eq. \ref{ln}. Eq. \ref{ln} is often used in the
literature, but it is only valid for $\gamma \gg MR$. This
relation does not apply to, e.g., the Higgs boson production
($M_{H^0} \sim 100$ GeV), as considered in ref.
\cite{Pa89,Grab89,RN00}.

\begin{figure}[t]
\centerline{\psfig{file=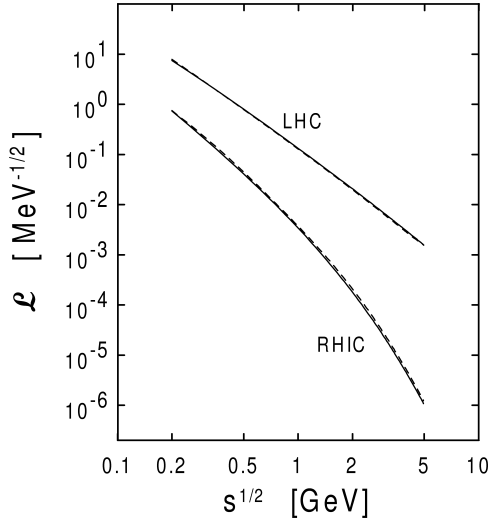,width=5cm}}
\vspace*{8pt}\caption{ Two-photon luminosities (see definition in
eq. \ref{sigfin}) at RHIC and LHC. Dashed lines include a
geometric correction. } \label{f3}
\end{figure}

For quantitative predictions one should use a more realistic form
factor. The Woods-Saxon distribution, with central density
$\rho_0$, size $R$, and diffuseness $a$ gives a good description
of the densities of the nuclei involved in the calculation. For
$Au$ we use $R = 6.38$ fm, and $a = 0.535$ fm, with $\rho_0$
normalized so that $\int d^3 r \rho (r) = 197$. For $Pb$ the
appropriate numbers are 6.63 fm, 0.549 fm, and 208, respectively
\cite{Ja74}.  In table 2  we show the cross sections for the
production of $C=even$ mesons at RHIC ($Au+Au$) and LHC ($Pb+Pb$)
using the formalism described above.

\small
\begin{table}[t]
\caption{Cross sections for two-photon production of ($C=even$)
mesons at RHIC ($Au+Au$) and at LHC ($Pb+Pb$).}
\begin{center}
\begin{tabular}{|l|l|l|l|l|l|l|l|} \hline meson     & mass&
$\Gamma_{\gamma\gamma}$ & $\sigma^{(+)}$  &${\cal
N}_{2\gamma}^{RHIC}/10^3$& ${\cal N}_{2\gamma}^{LHC}/10^7$
&$\sigma^{RHIC}$  & $\sigma^{LHC}$ \\ \hline
 &[MeV] &[keV]& [nb]&&&[$\mu$b]&[mb]\\ \hline
 $\pi_0$   &  134  & $7.8 \times 10^{-3}$&99&  49     & 2.8 &  4940  & 28    \\
\hline
 $\eta$    &   547  & 0.46     &   86 &       12      & 1.8 &  1000  &  16 \\
\hline
 $\eta'$   &   958  & 4.2     &   147 &         5.1      & 1.4 &  746  &  21 \\
\hline
 $f_2 (1270)$& 1275 & 2.4     &  179 &          3.0      & 1.2 &  544  & 22 \\
\hline
 $a_2 (1320)$& 1318  & 1.0     & 67  &        2.9      & 1.1 &  195  & 8.2   \\
\hline
 $\eta_c$  &   2981 & 7.5     &  8.7   &        0.38     & 0.7&  3.3 &  0.61  \\
\hline
 $\chi_{0c}$& 3415  & 3.3     &   2.6   &      0.24     & 0.63 &  0.63 & 0.16  \\
\hline
 $\chi_{2c}$&   3556  & 0.8     &  2.8 &       0.21     & 0.56&  0.59 & 0.15 \\
\hline
\end{tabular}
\end{center}
\end{table}
\normalsize

As pointed out in refs. \cite{BF90,CJ90}, one can improve the
(classical) calculation of the two-photon luminosities by
introducing a geometrical factor (the $\Theta$-function in ref.
\cite{BF90}), which affects the angular part of the integration
over impact parameters. This factor takes care of the position
where the meson is produced in the space surrounding the nuclei.
In the approach described here the form factors also introduce a
geometrical cutoff implying that the mesons cannot be produced
inside the nuclei. However, it is not easy to compare both
approaches directly, as one obtains a momentum representation of
the amplitudes when one performs the integration over impact
parameters to obtain eq. \ref{meson2}. But one can compare the
effects of geometry in both cases by using equation \ref{LLspec2}.
After performing the integral over $\omega$, one can rewrite it as
\begin{equation}
\sigma = \int ds  {\cal L} (s) \sigma_{\gamma\gamma} (s) \ ,
\label{sigfin}
\end{equation}
where $s = 4\omega_1\omega_2$ is the square of the center-of-mass
energy of the two photons, $\sigma(s)$ is given by eq. \ref{photon},
and ${\cal L}(s)$ is the ``photon-photon luminosity'', given by
\begin{eqnarray}
{\cal L}(s) &=& {1\over s} \ {Z^4\alpha^2\over \pi^2} \int
{d\omega\over \omega} \left[ \ln \left( 1+{\gamma^2 \over \omega^2
R^2} \right)- {1\over 1+\omega^2 R^2/\gamma^2}\right] \nonumber
\\
&\times&
\left[ \ln \left( 1+{16\gamma^2  \omega^2 \over s^2R^2}
\right)- {1\over 1+  s^2R^2/ 16\gamma^2  \omega^2  }\right]
\label{Luminosity} \ .
\end{eqnarray}

In figure \ref{f3} we compare the result obtained by eq.
\ref{Luminosity} and that of ref. \cite{BF90}. The luminosities
for RHIC ($Au+Au$) and for LHC ($Pb+Pb$) are presented. For RHIC
the difference between the two results can reach 10\% for very
large meson masses (e.g. the Higgs), but one notices that for the
LHC the two results are practically identical, the difference
being of the order of 3\%, or less, even for the Higgs production.
Thus, the improved version of eq. \ref{ln}, given by integrating
eq. \ref{LLspec2}, is accurate enough to describe meson production
by two-photon fusion. Other effects, like the interference between
the electromagnetic and the strong interaction production
mechanism in grazing collisions, must yield larger corrections to
the (non-disruptive) meson production cross sections than a more
elaborate description of geometrical effects.

One might think that the calculation above can be extended to the
three-photon fusion by using the equivalent photon approximation
that, as we have seen in this section, works so well for $C=even$
mesons. However, the introduction of a third photon leads to an
additional integration, which implies that at least two of the
exchanged photons cannot be treated as real ones. Nonetheless, the
results of this section paves the way to the calculation of
production of $C=odd$ mesons. Although the use of the projection
technique to systems composed of light quarks is questionable, we
have seen that it works, basically because of the relation
\ref{llM}, due to the inclusion of the nuclear form-factors.

\subsection{Three-photon Mechanism: Vector mesons}

Lets now consider the diagram of figure \ref{f1}(b), appropriate
for the fusion of three photons into a $C=odd$ particle. According
to the Feynman rules, the matrix element for it is given by
\begin{eqnarray}
{\cal M}_a &=& e^3 \bar{u}({P\over 2})\int{d^4q\over{(2\pi)^4}}
\int{d^4k\over{(2\pi)^4}} \not\!\! A^{(1)}({P\over 2}-q) {{\not\!
q + M/2}\over {q^2 - M^2/4}} \nonumber \\
&\times& \not\!\! A^{(2)}(q-k) {{\not\! k + M/2}\over {k^2 -
M^2/4}} \not\!\! A^{(2)}({P\over 2}+k)
 v({P\over 2}) \ .\label{M31}
\end{eqnarray}

There will be 12 diagrams like this. But the upper photon leg in
diagram of fig. \ref{f1}(b) can be treated as a real photon,
meaning that the equivalent photon approximation is valid for this
piece of the diagram.

\begin{figure}[t]
\centerline{\psfig{file=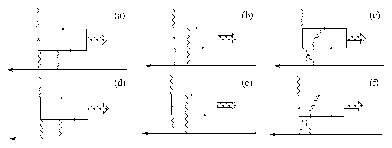,width=12cm}}
\vspace*{8pt}\caption{\small Feynman graphs for three-photon
fusion in ultra-peripheral collisions of relativistic heavy ions.
} \label{f2}
\end{figure}

A straightforward calculation yields
\begin{eqnarray}
{d\sigma \over d P_z} &=& 1024\pi \ \Big|\Psi(0)\Big|^2
(Z\alpha)^6 { 1\over M^3E} \ \int  { dq_{1t} \ q_{1t}^3\
\left[F(q_{1t}^2)\right]^2 \over \left(q_{1t}^2 +
\omega_2^2/\gamma^2 \right)^2} \nonumber \\
&\times& \int  {dq_{2t}\ q_{2t} \ \left[F(q_{2t}^2)\right]^2 \over
 \left[ q_{2t}^2+(2\omega_1-\omega_2)^2/\gamma^2\right]^2}
\left[ \int {dk_t \ k_t\ F(k_t^2) \over
\left(k_t^2+(\omega_1-\omega_2)^2/\gamma^2\right)}
\right]^2
\end{eqnarray}

We now use the relationship between $E$ and $P_z$ to $\omega_1$
and $\omega_2$ and get rid of the meson wavefunction at the
origin. The wavefunction $\Big|\Psi(0)\Big|^2$ cannot be related
to the $\gamma\gamma$ decay widths. But, vector mesons can decay
into $e^+e^-$ pairs. These decay widths are very well known
experimentally. Following a similar derivation as for the
$\gamma\gamma$-decay the $e^+e^-$ decay-width of the vector mesons
can be shown \cite{RW67} to be equal to $\Gamma_{e^+e^-}=16\pi
\alpha^2 \Big|\Psi(0)\Big|^2/3M^2 \ (3\cdot \sum_i Q_i^2)$.
Inserting these results in the above equation, the factor $(3\cdot
\sum_i Q_i^2)$ will cancel out for the same reason as explained in
section 3, and one gets
\begin{equation}
{d\sigma \over d\omega} =  \sigma^{(-)} \
\ {n(\omega)\over \omega}\ \ H (M,\omega)
\end{equation}
where
\begin{equation}
 \sigma^{(-)}=96\pi
{\Gamma_{e^+e^-}\over M^3} \ ,
\end{equation}
with $n(\omega)$ given by  \ref{epaf2} and
\begin{eqnarray}
H(M,\omega) &=& Z^4\alpha^3 M^2 \int {dq_{2t}\ q_{2t} \
\left[F(q_{2t}^2)\right]^2 \over \left[
q_{2t}^2+(M^2/2\omega-\omega)^2/\gamma^2\right]^2}
\nonumber  \\
&\times& \left[ \int {dk_t \ k_t\ F(k_t^2) \over
\left(k_t^2+(M^2/4\omega-\omega)^2/\gamma^2\right)^2}\right]^2 \
.
\label{three5}
\end{eqnarray}

The above formulas should also be valid for the production of the
ortho-positro\-nium in ultra-peripheral collisions of relativistic
heavy ions. For RHIC ($Au+Au$) one obtains $\sigma = 11.2$ mb,
while for the LHC ($Pb+Pb$) we get $\sigma = 35$ mb. These numbers
are also in good agreement with the results (in the Born
approximation) given in ref. \cite{Kot99}. When one includes
Coulomb corrections, as shown in ref. \cite{Kot99}, the cross
sections for orhto-positronium production is reduced by 40\% for
both RHIC and LHC. This is not considered in the present approach
as we are mainly interested in vector meson production for which
this effect should be smaller.

In table 3 we present the cross sections for the production of
vector mesons by means of the three-photon fusion process are
given.

\small
\begin{table}[t]
\caption{Cross sections for three-photon production of vector
($C=odd$) mesons at RHIC ($Au+Au$) and at LHC ($Pb+Pb$).}
\begin{center}
\begin{tabular}{|l|l|l|l|l|l|}
\hline meson     & mass [MeV]& $\Gamma_{e^+e^-}$ [keV]&
$\sigma^{(-)}$ [nb] &$\sigma^{RHIC}$ [nb]
& $\sigma^{LHC}$ [nb] \\
\hline
$\rho^0$ &  770  & 6.77 &1740 & 137   & 1801    \\
\hline
$\omega$ & 782  & 0.60     & 147 &  13 & 163 \\
\hline
$J/\psi$& 3097 & 5.26 & 21 & 31  & 423 \\
\hline
$\Psi$'& 3686 & 2.12 & 5 & 12 & 155 \\
\hline
\end{tabular}
\end{center}
\end{table}
\normalsize

One sees that the cross sections for the production of vector
mesons in ultra-peripheral collisions  of relativistic heavy ions
are small. They are not comparable to the production of vector
mesons in central collisions. In principle, one would expect that
the cross sections for three-photon production would scale as
$(Z\alpha)^3$, which is an extra $Z\alpha$ factor compared to the
two-photon fusion cross sections. However, the integral over the
additional photon momentum decreases the cross section by several
orders of magnitude.

In the next section we discuss the possibility to access
information on the gluon distribution in nuclei using the abundant
virtual photons in PHIC \cite{VB02}.

\section{One-photon Production Mechanism}

\subsection{Vector Mesons and Gluon Distributions}

One of main predictions of QCD is the transition from the
confined/chirally broken phase to the deconfined/chirally
symmetric state of quasi-free quarks and gluons, the so-called
quark-gluon plasma (QGP). Recently the heavy ion collisions have
provided strong evidence for the formation of a QGP \cite{sps},
with the first results of  RHIC marking the beginning of a
collider era in the experiments with relativistic heavy ions, as
well as the  era of detailed studies of  the characteristics of
the QGP. Currently, distinct models associated to different
assumptions describe reasonably the experimental data
\cite{eskqm}, with the main uncertainty present in these analysis
directly connected with the poor knowledge of the initial
conditions of the heavy ion collisions. Theoretically, the early
evolution of these nuclear collisions is governed by the dominant
role of gluons \cite{vni}, due to their large interaction
probability and the large gluonic component in the initial nuclear
wave functions. This leads to a ``hot gluon scenario'', in which
the large number of initially produced energetic partons create a
high temperature, high density plasma of predominantly hot gluons
and a considerably number of quarks. Such extreme condition is
expected to significantly influence QGP signals and should modify
the hard probes produced at early times of the heavy ion
collision. Consequently, a systematic measurement of the nuclear
gluon distribution is of fundamental interest in understanding of
the parton structure of nuclei and to determine the initial
conditions of the QGP. Other important motivation for the
determination of the nuclear gluon distributions is that the high
density effects expected to occur in the high energy limit of QCD
should be manifest in the modification of the gluon dynamics.

At small $x$ and/or large $A$ we expect the transition of the
regime described by the linear dynamics (DGLAP, BFKL) (for a
review, see e.g. ref. \cite{caldwell}), where only the parton
emissions are considered, to a new regime where the physical
process of recombination of partons becomes important in the
parton cascade and the evolution is given by a nonlinear evolution
equation. In this  regime    a Color Glass Condensate (CGC) is
expected to be formed \cite{mcl}, being characterized by the
limitation on the maximum phase-space parton density that can be
reached in the hadron/nuclear wavefunction (parton saturation) and
very high values of the QCD field strength $F_{\mu \nu} \approx
1/\sqrt{\alpha_s}$ \cite{mue}. In this case, the number of gluons
per unit phase space volume practically saturates and at large
densities grows only very slowly (logarithmically) as a function
of the energy \cite{vicsat}. This implies a large modification of
the gluon distribution if compared with the predictions of the
linear dynamics, which is amplified in nuclear processes
\cite{agl,mclgos,df2a,vicslope}.

Other medium effects are also expected to be present in the
nuclear gluon distribution at large values of $x$: the
antishadowing ($0.1 < x < 0.3$), the EMC effect ($0.3 < x < 0.7$)
and the Fermi motion ($x > 0.7$) \cite{arneodo,weise}. The
presence of these effects is induced from the experimental data
for the nuclear structure function which determines the behavior
of the nuclear quark distributions and the use of the momentum sum
rule as constraint. Experimentally, the behavior of the  nuclear
gluon distribution is indirectly determined in the lepton-nucleus
processes in a small kinematic range of the fixed target
experiments, with the behavior at small $x$ (high energy)
completely undefined. This situation should be improved in the
future with the electron-nucleus colliders at HERA and RHIC
\cite{heraa,raju}, which probably could determine whether parton
distributions saturate. However, until these colliders become
reality we need to consider alternative searches in the current
accelerators which allow us to constraint the nuclear gluon
distribution. In this section we analyze the possibility of using
peripheral heavy ion collisions as a photonuclear collider  and
therefore to determine the behavior of the gluon distribution.

A photon stemming from the electromagnetic field of one of the two
colliding nuclei can penetrate into the other nucleus and interact
with one or more of its hadrons,   giving rise to photon-nucleus
collisions to an energy region hitherto unexplored experimentally.
For example, the interaction of quasireal photons with protons has
been studied extensively at the electron-proton collider at HERA,
with $\sqrt{s} = 300$ GeV. The obtained $\gamma p$ center of mass
energies extends up to $W_{\gamma p} \approx 200$ GeV, an order of
magnitude larger than those reached by fixed target experiments.
Due to the larger number of photons coming from one of the
colliding nuclei in heavy ion collisions similar and more detailed
studies will be possible in these collisions, with $W_{\gamma N}$
reaching 950 GeV for the Large Hadron Collider (LHC) operating in
its heavy ion mode.

\begin{figure}[t]
\centerline{\psfig{file=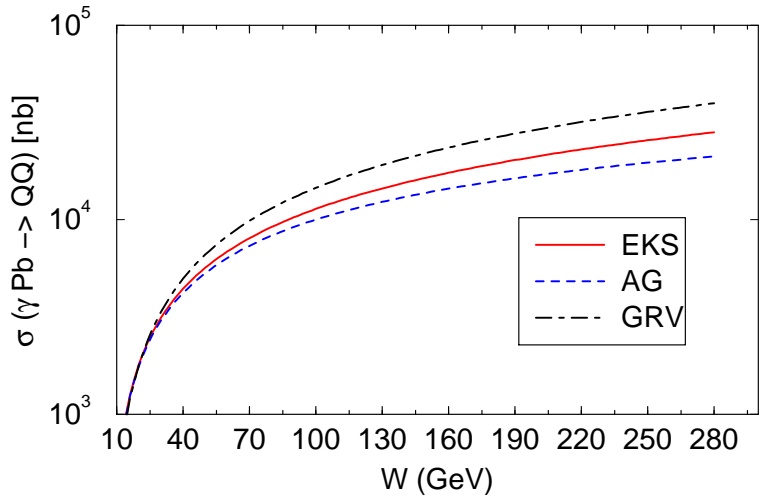,width=100mm}} \caption{Energy
dependence of the photoproduction of heavy quarks for distinct
nuclear gluon distributions ($A=208$).} \label{fig1}
\end{figure}

When a  very hard photon from one equivalent swarm of photons
penetrates the other nucleus it is able to resolve the partonic
structure of the nucleus and to interact with the quarks and
gluons. One of the basic process which can occur in the high
energy limit is the photon-gluon fusion leading  to the production
of a quark pair. The main characteristic of this process is that
the cross section is directly proportional to the nuclear gluon
distribution. The analysis of this process in peripheral heavy ion
collisions has been proposed many years ago \cite{perihe1} and
improved in the Refs. \cite{perihe2,perihe3} (for a review, see
ref. \cite{perihe4}). Here we reanalyze the charm photoproduction
as a way to estimate the medium effects in $xG_A$ in the full
kinematic region. We consider  as input three distinct
parameterizations of the nuclear gluon distribution. First,  we
consider the possibility that the nuclear gluon distribution is
not modified by medium effects, i.e., $xG_A (x,Q^2) = A \times
xG_N (x,Q^2)$, with the nucleon gluon distribution ($xG_N$) given
by the GRV parameterization \cite{grv95}. Moreover, we consider
that $xG_A (x,Q^2) = R_G \times xG_N (x,Q^2)$, where $R_G$
parameterizes the medium effects as proposed by Eskola, Kolhinen
and Salgado (EKS) \cite{eks}. The main shortcoming of these
parameterizations is that these are based on the DGLAP evolution
equations which are not valid in the small $x$ regime, where the
parton saturation effects should be considered. In order to
analyze the sensitivity of peripheral heavy ion collisions to
these effects we consider as input the parameterization proposed
by Ayala Filho and Gon\c calves (AG) \cite{ag} which improves the
EKS parameterization to include the high density effects.

One shortcoming of the analysis of  photoproduction of heavy
quarks in peripheral heavy ion collisions to constraint the
nuclear gluon distribution is the linear dependence of the cross
section with this distribution. This implies that only
experimental data with a large statistics and small error will
allow to discriminate the medium effects in the nuclear gluon
distribution. Consequently, it is very important to analyze other
possible processes which have a stronger dependence in $xG_A$.
Here we propose the study of the elastic photoproduction of vector
mesons in peripheral heavy ion collisions as a probe of the
behavior of the nuclear gluon distribution. This process has been
largely studied in $ep$ collisions at HERA, with the perturbative
QCD predictions describing successfully the experimental data
\cite{caldwell}, considering a quadratic dependence of the cross
section with the nucleon gluon distribution.

\subsection{Photoproduction of Heavy Quarks}

At high energies the dominant process occurring when the photon
probes the structure of the nucleus is the photon-gluon fusion
producing a quark pair. For heavy quarks the photoproduction can
be described using perturbative QCD, with the cross section given
in terms of the convolution between the elementary cross section
for the sub-process $\gamma g \rightarrow Q \overline{Q}$ and the
probability of finding a gluon inside the nucleus, i.e., the
nuclear gluon distribution. Basically, the  cross section for
$c\overline{c}$ photoproduction is given by
\begin{eqnarray}
\sigma_{\gamma g\rightarrow c\overline{c}}\, (s) &=&
\int_{2m_c}^{\sqrt s}dM_{c\overline{c}}
    \frac{d\sigma_{c\overline{c}}}{dM_{c\overline{c}}}\, g_A(x,\mu)\,\,,
    \label{sigpho}
\end{eqnarray}
where $d\sigma_{c\overline{c}}/dM_{c\overline{c}}$ is calculable
perturbatively, $M_{c\overline{c}}$ is the invariant mass of the
$c\overline{c}$ pair with $M^2_{c\overline{c}}={\hat{s}}=xs$, $s$
is the squared CM energy of the $\gamma A$ system, $g_A(x, \mu)$
is the gluon density inside the nuclear medium, $\mu$ is the
factorization scale ($\mu = \sqrt{M^2_{c\overline{c}}}$), and
$m_c$ is the charm quark mass (we assume that $m_c = 1.45$ GeV).
Moreover, the differential cross section is \cite{Gluck78}
\begin{eqnarray}
\frac{d\sigma_{\gamma g\rightarrow c\overline{c}}}
{dM_{c\overline{c}}} = \frac {4\pi\alpha\alpha_se_c^2}
{M^2_{c\overline{c}}} \Big[
(1+\epsilon+\frac{1}{2}\epsilon^2)\ln(\frac{1+\sqrt{1-\epsilon}}
{1-\sqrt{1-\epsilon}})  -(1+\epsilon) \sqrt{1-\epsilon}\Big] \,\,,
\label{integ}
\end{eqnarray}
where $e_c$ is the charm charge and
$\epsilon=4m_c^2/M_{c\overline{c}}^2$. From the above expression,
we verify that the cross section is directly proportional to the
nuclear gluon distribution, which implies the possibility  to
constraint its behavior from experimental results for
photoproduction of heavy quarks.

Throughout this section we use the Born expression for the
elementary photon-gluon cross section [eq. (\ref{integ})]. QCD
corrections to the Born cross section will not be considered here,
although these corrections modify the normalization of the cross
section by a factor of two. This is justified by the fact that we
are interested in the relative difference between the predictions
of the distinct nuclear gluon distributions, which should be not
modified by the next-to-leading-order corrections.

\begin{figure}[t]
\centerline{\psfig{file=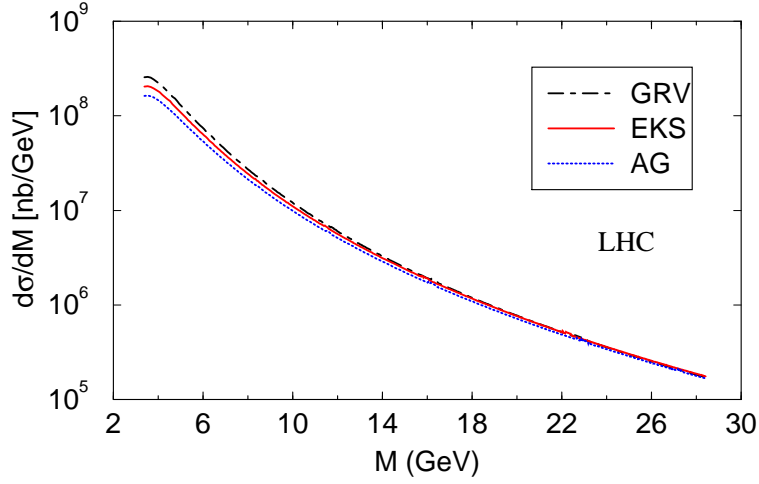,width=100mm}}
\caption{Differential cross section for $c\overline{c}$ production
versus the invariant mass $M=M_{c\overline{c}}$. The photon-gluon
fusion in the heavy-ion collision system $^{208}$Pb + $^{208}$Pb
at LHC energy is considered.} \label{fig2}
\end{figure}

In Fig. \ref{fig1} we present the energy dependence of the
photoproduction cross section. We focus  our discussion on charm
photoproduction instead of bottom photoproduction, since in this
process smaller values of $x$ are probed.  One verifies that
different nuclear gluon distributions imply distinct behaviors for
the cross section, with the difference between the predictions
increasing with the energy. This is associated to the fact that at
high energies we are probing the small $x$ behavior of $xG_A$,
since $x \propto M_{c\overline{c}}/s$, where $M_{c\overline{c}}$
is the invariant mass of the photon-gluon system. Currently, only
the region of small center of mass energy has been analyzed by the
fixed target electron-nucleus experiments, not allowing a good
constraint on medium effects present in the nuclear gluon
distribution. Such situation should be improved in the future with
electron-nucleus colliders at HERA and RHIC \cite{arneodo,raju}.

Another possibility to study photoproduction of heavy quarks at
large center of mass energies is to consider  peripheral heavy ion
collisions \cite{perihe1,perihe2,perihe3}. In this process the
large number of photons coming from one of the colliding nuclei in
heavy ion collisions will allow to study photoproduction, with
$W_{\gamma N}$ reaching  950 GeV for the LHC. To determine the
photoproduction of heavy quarks in peripheral heavy ion collisions
the elementary photon-gluon cross section has to be convoluted
with the photon energy distribution and the gluon distribution
inside the nucleus:
\begin{eqnarray}
\sigma (AA \rightarrow XXQ\overline{Q}) = n(\omega) \otimes
\sigma_{\gamma g \rightarrow Q\overline{Q}} \otimes
xG_A(x,Q^2)\,\,. \label{direct}
\end{eqnarray}

It is interesting to determine the values of $x$ which will be
probe in peripheral heavy ion collisions. The Bjorken $x$ variable
is given by $x = (M/2p) e^{-y}$, where $M$ is the invariant mass
of the photon-gluon system and $y$ the center of momentum
rapidity. For Pb + Pb collisions at LHC energies the nucleon
momentum is equal to $p=3000$ GeV; hence $x = (M/6000 \, {\rm
GeV}) e^{-y}$. Therefore, the region of small mass and large
rapidities probes directly the small $x$ behavior of the nuclear
gluon distribution. This demonstrates that peripheral heavy ion
collisions at LHC represents a very good tool to determine the
behavior of the gluon distribution in a nuclear medium, and in
particular the  low $x$ regime. Conversely, the region of large
mass and small rapidities is directly associated to the region
where the EMC and antishadowing effects are expected to be
present. Similarly, for RHIC energies  ($p = 100$ GeV) the cross
section will probe the region of medium and large values of $x$
($x > 10^{-2}$). For this kinematical region, the EKS and AG
parameterizations are identical, which implies that the
photoproduction  of heavy quarks in peripheral heavy ion
collisions at RHIC does not allows to constraint the high density
effects. However, the study of this process at RHIC will be very
interesting to determine the presence or not of the antishadowing
and EMC effect in the nuclear gluon distribution.

\begin{figure}[t]
\centerline{\psfig{file=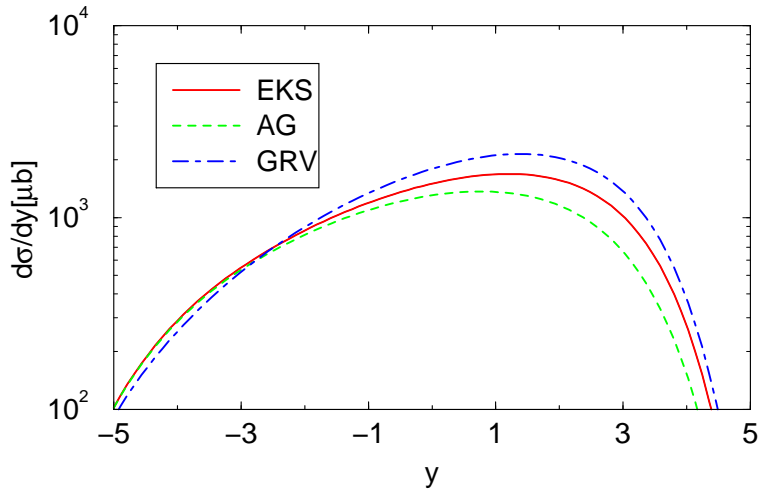,width=100mm}}
\caption{Rapidity distribution for the photoproduction of charm
quarks  in $^{208}$Pb + $^{208}$Pb collisions at  LHC.}
\label{fig3x}
\end{figure}

 Fig. \ref{fig2} shows the mass
distribution in photoproduction of charm quarks in peripheral
heavy ion collisions. We can see that the main difference between
the predictions occur at small values of $M$, which is associated
to the small $x$ region. Basically, we have that the predictions
of the EKS parameterization are a factor 1.25 larger than the AG
prediction in this region, while the prediction of Gl\"{u}ck, Reya
and Vogt (GRV) is a factor 2.4 larger. This result is consistent
with the fact that the main differences between the
parameterizations of the nuclear gluon distribution occur at small
$x$ \cite{ag}. The difference between the predictions diminishes
with the growth of the invariant mass, which implies that this
distribution is not a good quantity to estimate the nuclear
effects for medium and large $x$.

A better distribution to discriminate the behavior of the nuclear
gluon distribution is the rapidity distribution, which is directly
associated to the Bjorken $x$ variable, as discussed above. The
rapidity distribution is calculated considering that $d\sigma/dy =
\omega d\sigma/d\omega$. In Fig. \ref{fig3x} we show the rapidity
distribution considering the three parameterizations of $xG_A$ as
input. We have that the region of small rapidities probes the
region of large $x$, while for large $y$ we directly discriminate
the different predictions of $xG_A$ for small $x$. These results
are coherent with this picture: as at large $x$ the EKS and AG
predictions are identical, this region allows to estimate $R_G =
xG_A/(AxG_N)$ in the region of the antishadowing and EMC effects;
at large $y$ the large difference between the parameterizations
implies large modifications in the rapidity distribution, which
should allow a clean  experimental analysis.

\begin{figure}[t]
\centerline{\psfig{file=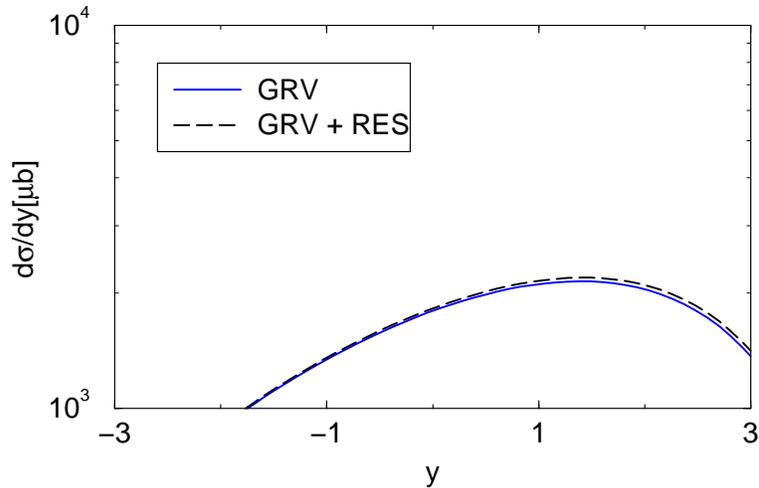,width=100mm}}
\caption{Rapidity distribution for the photoproduction of charm
quarks  in $^{208}$Pb + $^{208}$Pb collisions at  LHC with
(GRV+RES) and without (GRV) the inclusion of the resolved
contribution.} \label{fig3b}
\end{figure}

A comment is in order here. In hard photon-hadron interactions the
photon can behave as a pointlike particle in the so-called direct
photon processes or it can act as a source of partons, which then
scatter against partons in the hadron, in the resolved photon
processes (for a recent review see ref. \cite{nisius}). Resolved
interactions stem from the photon fluctuation to a quark-antiquark
state or a more complex partonic state, which are embedded in the
definition of the photon structure functions. Recently, the
process of jet production in  photoproduction has been search of
studies of the partonic structure of the photon (See e.g.
\cite{caldwell}), and the contribution of the resolved photon for
the photoproduction of charm has been estimated \cite{frixione}.
Basically, these studies shown that the partonic structure of the
photon is particularly important in some kinematic regions (for
example, the region of large transverse momentum of the charm pair
\cite{frixione}). Consequently, it is important to analyze if this
contribution modifies, for instance, the results for the rapidity
distribution, which is strongly dependent of  nuclear gluon
distribution. To leading order, beyond the process of photon-gluon
fusion considered above, charm production can occur also in
resolved photon interactions, mainly through the process $gg
\rightarrow c\overline{c}$. Therefore, one needs to add in eq.
(\ref{direct}) the resolved contribution  given by
\begin{eqnarray}
\sigma_{res} (AA \rightarrow XXQ\overline{Q}) = n(\omega) \otimes
x_{\gamma}G_{\gamma}(x_{\gamma},Q^2) \otimes \sigma_{g g
\rightarrow Q\overline{Q}} \otimes xG_A(x,Q^2)\,\,,
\label{resolved}
\end{eqnarray}
where $x_{\gamma}$  denote the fraction of the photon momentum
carried by its  gluon component $x_{\gamma} G_{\gamma}$  and
$\sigma_{g g \rightarrow Q\overline{Q}}$ the heavy quark
production cross section first calculated in ref.
\cite{combridge}. Since the resolved contribution should be the
same for the three nuclear parton distributions, Fig. \ref{fig3b}
only shows the results obtained using the  GRV parameterization
for the nucleon. For the photon distribution we use the GRV photon
parameterization \cite{grvphoton}, which predicts a strong growth
of the photon gluon distribution at small $x_{\gamma}$. We can see
that though this contribution is important in the photoproduction
of heavy quarks, as shown in ref. \cite{frixione}, it is small in
the rapidity distribution of charm quarks produced in peripheral
heavy ion collisions. Therefore, the inclusion of the resolved
component of the photon does not is not relevant for the use of
this process as a probe of the nuclear gluon distribution.

It is important to salient that the potentiality of the
photoproduction of quarks to probe the high density effects have
been recently emphasized in ref. \cite{gelis}, where the color
glass condensate formalism was used to estimate the cross section
and transverse momentum spectrum. The authors have verified that
the cross section is sensitive to the saturation scale which
characterizes the colored glass. The results abovecorroborate the
conclusion that this process is sensitive to the high density
effects and the agreement between the predictions is expected in
the kinematic  region in which the transverse momentum of the pair
$k_t$ is larger than the saturation scale $Q_s$. For $k_t < Q_s$
the collinear factorization to calculate the cross sections must
be generalized, similarly to ref. \cite{gelis}.

\subsection{Elastic Photoproduction of Vector Mesons}

The production of vector mesons at HERA has become a rich field of
experimental and theoretical research [for a recent review see,
e.g. ref. \cite{revjpsi}], mainly related with the question of
whether perturbative QCD (pQCD) can provide an accurate
description of  the elastic photoproduction processes. At high
energies the elastic photoproduction of vector mesons is a
two-stage process: at  first the photon fluctuates into the vector
meson which then interacts with the target. For light vector
mesons the latter process occurs similarly to the soft
hadron-hadron interactions and can be interpreted within Regge
phenomenology \cite{dl}. However, at large mass of the vector
meson, for instance the mass of the $J/\Psi$ meson, the process is
hard and pQCD can be applied \cite{brod}. In this case the
lifetime of the quark-antiquark fluctuation is large compared with
the typical interaction time scale and the formation of the vector
meson only occur after the interaction with the target. In pQCD
the interaction of the $q\overline{q}$ pair is described by the
exchange of a color singlet system of gluons (two gluons to
leading order) and, contrary to the Regge approach,  a steep rise
of the vector meson cross section is predicted, driven by the
gluon distribution in the proton. Measurements of the elastic
photoproduction of $J/\Psi$ mesons in $ep$ processes has been
obtained by the H1 and ZEUS collaborations for values of center of
mass energy below 300 GeV, demonstrating the steep rise of  the
cross section predicted by pQCD. This result motivates the
extension of the  pQCD approach used in electron-proton collisions
to photonuclear processes.

\begin{figure}[t]
\centerline{\psfig{file=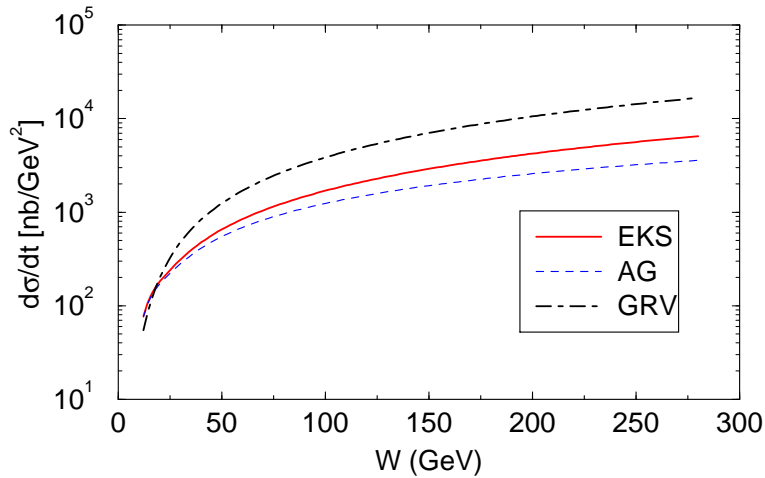,width=100mm}} \caption{Energy
dependence of the elastic photoproduction of $J/\Psi$ for distinct
nuclear gluon distributions ($A=208$).} \label{fig4x}
\end{figure}

The procedure for calculating the forward differential cross
section for photoproduction of a heavy vector meson in the color
dipole approximation is straightforward. The calculation was
performed some years ago to leading logarithmic approximation,
assuming the produced vector-meson quarkonium system to be
nonrelativistic \cite{brod} and improved in distinct aspects
\cite{fran}. To lowest order the $\gamma A \rightarrow J/\Psi A$
amplitude can be factorized into the product of the $\gamma
\rightarrow c \overline{c}$ transition, the scattering of the
$c\overline{c}$ system on the nucleus via (colorless) two-gluon
exchange, and finally the formation of the $J/\Psi$ from the
outgoing $c\overline{c}$ pair.  The heavy meson mass $M_{J/\Psi}$
ensures that pQCD can be applied to photoproduction. The
contribution of pQCD to the imaginary part of the $t=0$
differential cross section of photoproduction of heavy vector
mesons is given by \cite{brod}
\begin{eqnarray}
\frac{d\sigma(\gamma A \rightarrow J/\Psi A)}{dt}|_{t=0} =
\frac{\pi^3 \Gamma_{ee} M_{J/\Psi}^3}{48 \alpha}
\frac{\alpha_s^2(\overline{Q}^2)}{\overline{Q}^8} \times
[xG_A(x,\overline{Q}^2)]^2\,\,, \label{sigela}
\end{eqnarray}
where $xG_A$ is the nuclear gluon distribution, $x =
4\overline{Q}^2/W^2$ with $W$ the center of mass energy and
$\overline{Q}^2 = M_{J/\Psi}^2/4$. Moreover, $\Gamma_{ee}$ is the
leptonic decay width of the vector meson. The total cross section
is obtained by integrating over the momentum transfer $t$,
\begin{eqnarray}
\sigma(\gamma A \rightarrow J/\Psi A) = \frac{d\sigma(\gamma A
\rightarrow J/\Psi A)}{dt}|_{t=0} \, \int_{t_{min}}^{\infty} dt \,
|F(t)|^2 \,\,\,, \label{photonuc}
\end{eqnarray}
where $t_{min} = (M_{J/\Psi}^2/2\omega)^2$ and $F(t)=\int d^3r \
\rho(r) \ \exp (i{\bf q}\cdot {\bf r})$ is the nuclear form factor
for the matter distribution.

\begin{figure}[t]
\centerline{\psfig{file=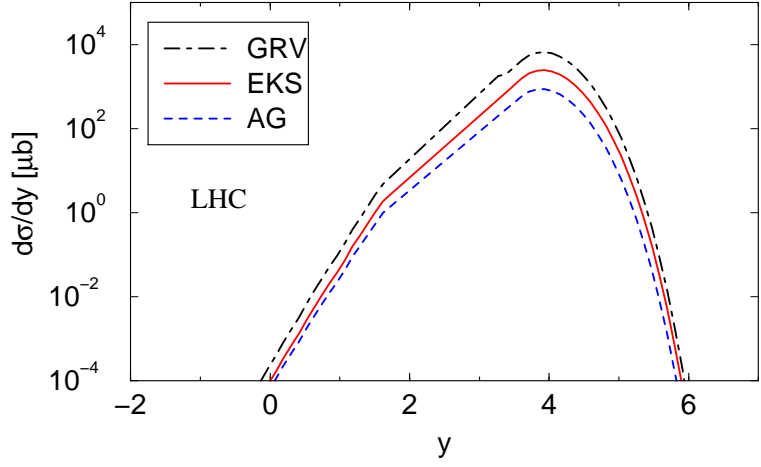,width=100mm}} \caption{The
rapidity distribution for elastic photoproduction of $J/\Psi$ at
LHC considering  distinct nuclear gluon distributions ($A=208$).}
\label{fig6}
\end{figure}

A comment is in order here. Although some improvements of the
expression (\ref{sigela})  have been proposed in the literature
\cite{fran}, these modifications do not change the quadratic
dependence on $xG_A$.

The main characteristic of the elastic photoproduction of vector
mesons is the quadratic dependence on the gluon distribution,
which makes it an excellent probe of the behavior of this
distribution. In Fig. \ref{fig4x} we show the energy dependence of
the differential cross section [eq. (\ref{sigela})], considering
the three distinct nuclear gluon distributions discussed above.
One obtains larger differences between the predictions than
obtained in photoproduction of heavy quarks, mainly at large
values of energy. This result motivates experimental analysis of
elastic $J/\Psi$ photoproduction  in photonuclear processes at
high energies. Although the future electron ion colliders (HERA-A
and eRHIC) should probe this kinematic region, here we show that
this can also be done with peripheral heavy ion collisions.

Following similar steps used in photoproduction of heavy quarks,
photons coming from one of the colliding nuclei may interact with
the  other. For elastic photoproduction of $J/\Psi$ one can
consider that this photon decays into a $c\overline{c}$ pair which
interacts with the nuclei by the two gluon exchange. After the
interaction, this pair becomes the heavy quarkonium state.
Consequently, the total  cross section for $J/\Psi$ production in
peripheral heavy ion collisions is obtained by integrating the
photonuclear cross section [eq. (\ref{photonuc})] over the photon
spectrum, resulting
\begin{eqnarray}
\sigma(AA\rightarrow AAJ/\Psi) = \int \frac{d\omega}{\omega} \
n(\omega) \frac{d\sigma(\gamma A \rightarrow J/\Psi A)}{dt}|_{t=0}
\, \int_{t_{min}}^{\infty} dt \, |F(t)|^2 \,\,.
\end{eqnarray}

\small
\begin{table}[t]
\caption{ The total cross section $\sigma(AA \rightarrow
AAJ/\Psi)$ for different nuclear gluon distributions. Results for
LHC.}
\begin{center}
\begin{tabular}{|l|l|} \hline
Gluon Distribution  & LHC \\
\hline GRV &  $6.584$ mb
\\\hline EKS &  2.452 mb \\
\hline AG &  0.893 mb \\
\hline
\end{tabular}
\end{center}
\end{table}
\normalsize

Table 4 presents the total cross section considering as input the
distinct nuclear gluon distributions and LHC energies. Although
these numbers will be modified by the inclusion of higher order
corrections for the cross section \cite{fran}, the difference
between the predictions should not be altered.  One verifies that
due to the large number of equivalent photons and the large center
of mass energies of the photon-nucleus system, the cross section
for this process is large, which allows an experimental
verification of these predictions. Also, in peripheral heavy ion
collisions the multiplicity is small what might simplify the
experimental analysis. Moreover, the difference between the
predictions is significant. For RHIC energies, the cross section
for this process is small and probably an experimental
determination will be very hard.

\begin{figure}[t]
\centerline{\psfig{file=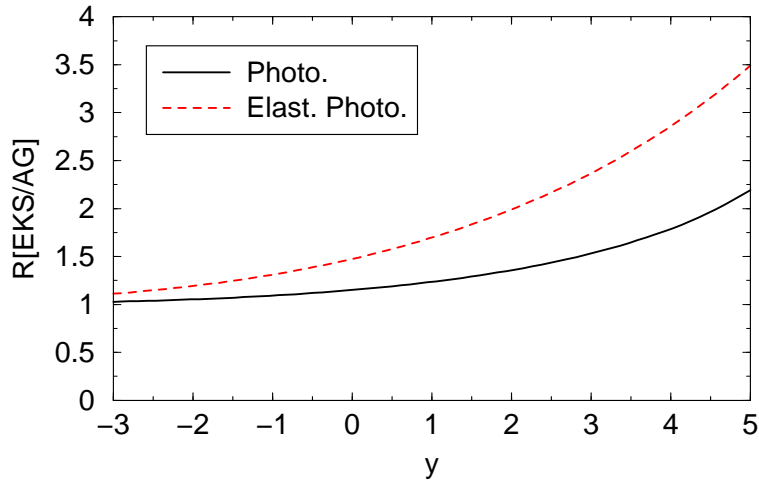,width=100mm}}
\caption{Rapidity behavior of the ratio between the EKS and AG
predictions for photoproduction of charm quarks and elastic
photoproduction of $J/\Psi$.} \label{fig7}
\end{figure}

In fig.  \ref{fig6} the rapidity distribution for $J/\Psi$
production at  LHC energies is shown. In this case, the final
state rapidity is determined by
\begin{eqnarray}
y = \frac{1}{2}\ln\frac{\omega}{\sqrt{|t_{min}|}} = \ln \frac{ 2
\omega}{M_{J/\Psi}} \,\,.
\end{eqnarray}
Similarly to the heavy quark photoproduction, the large $y$ region
probes the behavior of $xG_A$ at small $x$, while the region of
small $y$ probes medium values of $x$. One concludes that the
rapidity distribution for  elastic production of $J/\Psi$ at RHIC
allows to discriminate between the GRV and EKS prediction, with
the AG prediction being almost identical to the latter. For LHC
one finds a large difference between the distributions, mainly in
magnitude, which  will allow to estimate the magnitude of the EMC,
antishadowing and high density effects.

Finally, fig. \ref{fig7} compares the photoproduction of heavy
quarks and the elastic photoproduction of $J/\Psi$ in peripheral
heavy ion collisions as a possible process to constraint the
behavior of the nuclear gluon distribution. The rapidity
distribution of the ratio
\begin{eqnarray}
R[EKS/AG] \equiv \frac{d\sigma}{dy}[EKS]/\frac{d\sigma}{dy}[AG]
\,\,,
\end{eqnarray}
is shown where we consider the EKS and AG parameterizations as
inputs of the rapidity distributions. One confirms that the
analysis of the elastic photoproduction of $J/\Psi$ at medium and
large rapidities is a potential process to determinate the
presence and estimate the magnitude of the high density effects.

\section{Final Remarks}

The possibility to produce a Higgs boson via $\gamma\gamma$ fusion
was suggested in ref. \cite{Pa89,Grab89}. The cross sections for
LHC are of order of 1 nanobarn, about the same as for gluon-gluon
fusion. But, the two-photon processes can also produce $b\bar{b}$
pairs which create a large background for detecting the Higgs
boson. A good review of these topics was presented in ref.
\cite{BHT98}.

The excitation of a hadron in the field of a nucleus is another useful tool
to study the properties of hadrons. It has been used for example to obtain
the lifetime of the $\Sigma_0$ particle by measuring the (M1) excitation
cross section for the process $\gamma + \Lambda \rightarrow \Sigma_0$ \cite
{Dy77}. The vertex $\gamma \rightarrow 3\pi$ has been investigated \cite
{Ant87} in the reaction of pion pair production by pions in the nuclear
Coulomb field: $\pi^- +Z \rightarrow \pi^- +\pi^0 +Z$. Also, the $\pi^-$
polarizability has been studied in the reaction $\pi^- +Z \rightarrow \pi^-
+\gamma+Z$ \cite{Ant83}. Other unexplored possibilities includes the
excitation of a nucleon to a $\Delta$-particle in the field of a heavy
nucleus in order to disentangle the $M1$ and $E2$ parts of the excitation.

As for meson-production in PHIC there are several planned
experiments at RHIC, as well as for the future LHC
\cite{spencer3}. These machines were designed to study hadronic
processes. But, as have shown in this brief review, they can also
be used for studying very interesting phenomena induced in
peripheral collisions.

\section*{Acknowledgment(s)}

I have benefited from fruitful discussions with Sam Austin, A.
Baltz, G. Baur,  F. Gelis, V. Gon\c calves, F. Navarra, V. Serbo,
and S. Typel. This research was supported in part by the U.S.
National Science Foundation under Grants No. PHY-007091 and
PHY-00-70818.


\small
\begin{thebibliography}{99}

\bibitem{BB94}  C.A. Bertulani and G. Baur, Physics Today, March 1994,
p. 22.

\bibitem{Fe24}  E. Fermi, Z. Physik {\bf 29}, 315 (1924).

\bibitem{WW34} K. F. Weizsacker, Z.
Physik , 612 (1934); E. J. Williams, Phys. Rev. {\bf 45}, 729 (1934).

\bibitem{BB85}  C.A.Bertulani and G. Baur, Nucl. Phys. {\bf A442}
(1985) 73.

\bibitem{BB88}  C.A. Bertulani and G. Baur, Physics Reports {\bf 163}
(1988) 299.

\bibitem{BHTS02}
G. Baur, K. Hencken, D. Trautmann, S. Sadovsky, Y. Kharlov,
Phys. Rep. {\bf 364}, 359 (2002).

\bibitem{BCH93}  C.A. Bertulani, L.F. Canto and M.S. Hussein, Phys. Rep.
{\bf 226}, 282 (1993).

\bibitem{Glas98}
T. Glasmacher, Ann. Rev. Nucl. Part. Sci. {\bf 48}
(1998) 1.

\bibitem{Au98}  T. Aumann, P.F. Bortignon, H. Emling, Ann. Rev. Nucl.
Part. Sci. {\bf 48}, 282 (1998).

\bibitem{BP99} C.A. Bertulani and V.Yu. Ponomarev,
Phys. Reports {\bf 321}, 139 (1999).

\bibitem{AB95}  S.A. Austin and G.F. Bertsch, Scientific American,
 {\bf 272}, 62 (1995) 62

\bibitem{HJJ95}
P. G. Hansen, A. S. Jensen, and B. Jonson,{\it \ Ann. Rev.
Nucl. Part. Sci.} {\bf 45}, 591 (1995).

\bibitem{BBK92}  G. Baur, C.A. Bertulani and D.M. Kalassa, {\it Nucl. Phys.}
{\bf A550}, 527 (1992).

\bibitem{Ie93}
 K. Ieki, {\it et al.}, {\it Phys. Rev. Lett.} {\bf 70}
(1993) 730.

\bibitem{BBe93}
G.F. Bertsch and C.A. Bertulani, {\it Nucl. Phys.} {\bf A556}, 136
(1993); {\it Phys. Rev.} {\bf C49}, 2834 (1994).

\bibitem{HRB91}  M.S.Hussein, R.A.Rego and C.A. Bertulani, Phys. Reports
{\bf 201}, 279 (1991).

\bibitem{EBB95}
H. Esbensen, G.F. Bertsch and C.A. Bertulani, {\it Nucl. Phys}.
{\bf A581}, 107 (1995).

\bibitem{KYS96}
T. Kido, K. Yabana and Y. Suzuki, Phys.Rev. {\bf C53}, 2296
(1996).

\bibitem{Clayton}
D.D. Clayton, {\it Principles of Stellar Evolution and
Nucleosynthesis}, McGraw-Hill, New York, 1968.

\bibitem{Rolfs}
C. Rolfs and W.S. Rodney,
{\it Cauldrons in the Cosmos}, Chicago Press, Chicago, 1988.

\bibitem{BBH86}  G.Baur, C.A.Bertulani and H.Rebel, Nucl. Phys. \textbf{%
A458}, 188 (1986).

\bibitem{BCa92}
C.A.Bertulani and L.F.Canto, Nucl. Phys. {\bf A539}, 163 (1992).

\bibitem{CB95}
C.A. Bertulani, Z. Phys. {\bf A356}, 293 (1996).

\bibitem{NT99}
F.M. Nunes and I.J. Thompson, Phys. Rev. C59, 2652 (1999).

\bibitem{Mel99}  V. S. Melezhik and D. Baye, Phys. Rev. {\bf C59}, 3232
(1999).

\bibitem{paris99}  H. Utsunomia, Y. Tokimoto, T. Yamagata, M. Ohta, Y.
Aoki, K. Hirota, K. Ieki, Y. Iwata, K. Katori, S. Hamada, Y.-W. Lui, R. P.
Schmitt, S. Typel and G. Baur, Nucl. Phys. {\bf A654}, 928 (1999).


\bibitem{typwo99}  S. Typel, H. H. Wolter, Z. Naturforsch. {\bf 54 a}, 63 (1999).

\bibitem{Iw99}  N. Iwasa, et al., {\it Phys. Rev. Lett.} {\bf 83} (1999) 2910

\bibitem{DB94}
P. Descouvemont and D. Baye, Nucl. Phys. {\bf A567}, 341 (1994).

\bibitem{Dav01}
B. Davids, D. W. Anthony, T. Aumann, Sam M. Austin, T. Baumann, D.
Bazin, R. R. C. Clement, C. N. Davids, H. Esbensen, P. A. Lofy, T.
Nakamura, B. M. Sherrill, J. Yurkon, Phys.Rev.Lett. {\bf 86}, 2750
(2001).

\bibitem{Dav98}
B. Davids, D.W. Anthony, Sam M. Austin, D. Bazin, B. Blank, J.A.
Caggiano, M. Chartier, H. Esbensen, P. Hui, C.F. Powell, H.
Scheit, B.M. Sherrill, M. Steiner, P. Thirolf, Phys. Rev. Lett.
81, 2209 (1998).

\bibitem{Da01}
B. Davids, Sam M. Austin, D. Bazin, H. Esbensen, B. M. Sherrill, I. J. Thompson,
and J. A. Tostevin, Phys. Rev. {\bf C63}, 065806 (2001).

\bibitem{EB96}
H. Esbensen and G. Bertsch, Nucl. Phys. {\bf A600}, 37 (1996).

\bibitem{BR96}
G. Baur and H. Rebel, Annu. Rev. Nucl. Part. Sc. {\bf 46}, 321
(1996).

\bibitem{Kie93}
J. Kiener, A. Lefebre, P. Aguer, C.O. Bacri, R. Bimbot, G.
Bogaert, B. Borderie, F. Claper, A. Coc, D. Disdier, S. Fortier,
C. Grunberg, L. Klaus, I. Linck, G. Pasquier, M.F. Rivet, F.
StLaurent, C. Stephan, L. Tassangot and J.P. Thibaud, Nucl. Phys.
{\bf A552}, 66 (1993).

\bibitem{Kie91}
J. Kiener, H. J. Gils, H. Rebel, S. Zagromski, G. Gsottschneider,
N. Heide, H. Jelitto, J. Wentz and G. Baur, Phys. Rev. {\bf C44},
2195 (1991).

\bibitem{Mot91}
T. Motobayashi et al., Phys. Lett {\bf B264}, 259 (1991).

\bibitem{Lef95}
A. Lefebre et al., Nucl. Phys. {\bf A595}, 69 (1995).

\bibitem{Tat95}
V. Tatischeff, J. Kiener, P. Aguer, and A. Lefebvre Phys. Rev.
{\bf C51}, 2793 (1995).

\bibitem{BB86}
G. Baur and C.A. Bertulani, Phys. Lett. B 174 (1986) 23;
Nucl. Phys. A 482 (1988) 313; Phys. Rev. C 34 (1986) 1654;
Proc. Int. School of Heavy Ion Physics,
Erice, Italy, October 1986, Plenum Press, ed. by R.A. Broglia and G.F.
Bertsch, p. 331.

\bibitem{LB92}
W. Llope and P. Braun-Munzinger, Phys. Rev. C 45 (1992)
799; W. Llope, Ph. D. dissertation, SUNY at Stony Brook,
1992, and private communication.

\bibitem{Ri93}
J. Ritman, F.–D. Berg, W. Kühn, V. Metag, R. Novotny, M.
Notheisen, P. Paul, M. Pfeiffer, and O. Schwalb, H. Löhner and L.
Venema, A. Gobbi, N. Herrmann, K. D. Hildenbrand, J. Mösner, R. S.
Simon, K. Teh, J. P. Wessels, and T. Wienold, Phys. Rev. Lett.
{\bf 70}, 533 (1993).


\bibitem{Sc93}
R. Schmidt, Th. Blaich, Th. W. Elze, H. Emling, H. Freiesleben, K.
Grimm, W. Henning, R. Holzmann, J. G. Keller, H. Klingler, R.
Kulessa, J. V. Kratz, D. Lambrecht, J. S. Lange, Y. Leifels, E.
Lubkiewicz, E. F. Moore, E. Wajda, W. Prokopowicz, Ch. Schütter,
H. Spies, K. Stelzer, J. Stroth, W. Walus, H. J. Wollersheim, M.
Zinser, E. Zude, Phys. Rev. Lett. {\bf 70}, 1767 (1993).


\bibitem{Au93}
T. Aumann, J. V. Kratz, and E. Stiel, K. Sümmerer, W. Brüchle, M.
Schädel, and G. Wirth, M. Fauerbach  and J. Hill, Phys. Rev. {\bf
C47}, 1728 (1993).

\bibitem{Bor96}
K. Boretzky, J. Stroth, E. Wajda, T. Aumann, Th. Blaich, J. Cub,
Th. W. Elze, H. Emling, W. Henning, R. Holzmann, H. Klingler, R.
Kulessa, J. V. Kratz, D. Lambrecht, Y. Leifels, E. Lubkiewicz, K.
Stelzer, W. Walus, M. Zinser, E. Zude, Phys. Lett. {\bf B384}, 30
(1996 ).

\bibitem{Grun99}
A. Grünschloss, K. Boretzky, T. Aumann , C. A. Bertulani, J. Cub,
W. Dostal, B. Eberlein, Th. W. Elze, H. Emling, J. Holeczek, R.
Holzmann, M. Kaspar, J. V. Kratz, R. Kulessa, Y. Leifels, A.
Leistenschneider, E. Lubkiewicz, S. Mordechai, I. Peter, P.
Reiter, M. Rejmund, H. Simon, K. Stelzer, A. Surowiec, K.
Sümmerer, J. Stroth, E. Wajda, W. Walus, S. Wan, H. J.
Wollersheim, Phys. Rev. {\bf C60}, 051601 (1999).

\bibitem{Bor99}
K. Boretzky, T. Aumann, J. Cub, Th. W. Elze , H. Emling, A.
Grünschloss, R. Holzmann, S. Ilievski, R. Kulessa, J. V. Kratz, Y.
Leifels, A. Leistenschneider, E. Lubkiewicz, J. Stroth, E. Wajda,
W. Walus, Nucl. Phys. {\bf A649}, 235c (1999).

\bibitem{Ili01}
S. Ilievski, T. Aumann, K. Boretzky, J. Cub, W. Dostal, B.
Eberlein, Th. W. Elze, H. Emling, A. Grünschloss, J. Holeczek, R.
Holzmann, C. Kozhuharov, J. V. Kratz, R. Kulessa, Y. Leifels, A.
Leistenschneider, E. Lubkiewicz, T. Ohtsuki, P. Reiter, H. Simon,
K. Stelzer, J. Stroth, A. Surowiec, E. Wajda, W. Walus , Nucl.
Phys. {\bf A687} , 178c (2001).

\bibitem{Bert96}
C. A. Bertulani, L. F. Canto, M. S. Hussein, and A. F. R. de
Toledo Piza, Phys. Rev. {\bf C53}, 334 (1996).

\bibitem{HTV99}
M. S. Hussein, A. F. R. de Toledo Piza, and O. K. Vorov Phys. Rev.
{\bf C59}, R1242 (1999).

\bibitem{Tol99}
A. F. R. de Toledo Piza, M. S. Hussein, B. V. Carlson, C. A.
Bertulani, L. F. Canto, and S. Cruz-Barrios Phys. Rev. {\bf C59},
3093 (1999); B. V. Carlson, L. F. Canto, S. Cruz-Barrios, M. S.
Hussein, and A. F. R. de Toledo Piza Phys. Rev. {\bf C59}, 2689
(1999); B. V. Carlson and M. S. Hussein Phys. Rev. {\bf C59},
R2343 (1999); B. V. Carlson, M. S. Hussein, A. F. R. de Toledo
Piza, and L. F. Canto Phys. Rev. {\bf C60}, 014604 (1999).

\bibitem{Pau01}
D. T. de Paula, T. Aumann, L. F. Canto, B. V. Carlson, H. Emling,
and M. S. Hussein, Phys. Rev. {\bf C64}, 064605 (2001).

\bibitem{Psh01}
I.A. Pshenichnov, J.P. Bondorf, I.N. Mishustin, A. Ventura and S.
Masetti, Phys.Rev. C64,  024903 (2001).

\bibitem{seb1}  A.J. Baltz, M.J. Rhoades-Brown, and J. Weneser, Phys. Rev
{\bf E54}, 4233 (1996); A.J. Baltz, C. Chasman and S.N. White,
Nucl. Instrum. Meth, {\bf A417}, 1 (1998); S.N. White, Nucl.
Instrum. Meth. {\bf A409}, 618 (1998).

\bibitem{seb2}  M. Chiu, A. Denisov, E. Garcia, J. Katzy, A. Makeev, M. Murray,
and S. White, Phys. Rev. Lett. {\bf 89}, 012302 (2002).

\bibitem{Bal01}  A.J. Baltz, physics/0102045; see also, A.B. Voitkov, C.
M\"uller, and N. Gr\"un, Phys. Rev. {\bf A62}, 062701 (2000).

\bibitem{Kr98}  H.F. Krause {\it et al.}, {\it Phys. Rev. Lett.} {\bf 80}
(1998) 1190.

\bibitem{AB87}  R. Anholt and U. Becker, Phys. Rev. {\bf A36} (1987)
4628.

\bibitem{EM95}  J. Eichler and W. Meyerhof, {\it Relativistic Atomic
Collisions} (Academic Press, San Diego, 1995).

\bibitem{AG86}
R. Anholt and H. Gould, in A%
{\it Advances in Atomic and Molecular Physics} (Academic Press,
San Diego, 1986), p. 315.

\bibitem{Bec83}  U. Becker, N. Gr\"{u}n and W. Scheid, J. Phys. {\bf
B16} (1983) 1967; U. Becker, PhD. Thesis, Giessen (1986).

\bibitem{BS85}  C. Bottcher and M.R. Strayer, Phys. Rev. Lett. {\bf 54}
(1985) 669.

\bibitem{BB89}  G. Baur and C.A. Bertulani, Z. Phys. {\bf A330} (1988)
77; C.A. Bertulani and G. Baur, {\it Nucl. Phys}. {\bf A505} (1989) 835.

\bibitem{Gin92}  I.F. Ginzburg, G.L. Kotkin, S.I. Polityko and V.G. Serbo,
Phys. Rev. Lett. {\bf 68} (1992) 788; Phys. Lett. {\bf B286}
(1992) 392; Z. Phys. {\bf C60} (1993) 737.

\bibitem{serbo02} V. Serbo, private communication.

\bibitem{Kot99}
G.L. Kotkin, E.A. Kuraev, A. Schiller and V.G. Serbo, Phys. Rev.
{\bf C59} (1999) 2734

\bibitem{ERG00}  U. Eichmann, J. Reinhardt, and W. Greiner, Phys. Rev. A
{\bf 61}, 062710 (2000).

\bibitem{BGMP01}
A. Baltz, F. Gelis, L. McLerran and A. Peshier,
 Nucl. Phys. {\bf A695},  395 (2001).

\bibitem{Ba91}  A.J. Baltz, M.J. Rhoades-Brown, and J. Weneser, Phys. Rev. A
{\bf 44}, 5569 (1991); Phys. Rev. A {\bf 48}, 2002 (1993); Phys.
Rev. A {\bf 47}, 3444 (1993); Phys. Rev. A {\bf 50}, 4842 (1994);
A.J. Baltz, Phys. Rev. A {\bf 52}, 4970 (1995).

\bibitem{Ber01} C.A. Bertulani, Phys. Rev. {\bf A63}, 062706 (2001).

\bibitem{Vo00} A.B. Voitkiv, C. M\"uller, and N. Gr\"un, Phys. Rev.
{\bf A62}, 062701 (2000).

\bibitem{Fu33}  W.H. Furry and J.F. Carlson, Phys. Rev. {\bf 44}, 238
(1933).

\bibitem{LL34}
L.D. Landau and E.M. Lifshitz, Phys. Zs. Sowjet. {\bf 6}, 244
(1934).

\bibitem{Bha35}
H.J. Bhabha, Proc. R. Soc. London Ser. {\bf A152}, 559 (1935).

\bibitem{NTK35}
 Y. Nishina, S. Tomonaga and M. Kobayashi , Sci. Pap.
Phys. Chem. Res. {\bf 27}, 137 (1935).

\bibitem{RB91}  C. Bottcher and M.R. Strayer, Phys. Rev. {\bf
D39}, 1330
(1989); J. Phys. {\bf G16},  975 (1990); Phys. Lett. {\bf B23}%
7 (1990) 175.

\bibitem{Ba90}
G. Baur, Phys. Rev. {\bf D41}, 3535 (1990); Phys. Rev. {\bf A42},
5736 (1990).

\bibitem{RBW91}
M.J. Rhoades-Brown and J. Weneser, Phys. Rev. {\bf A4},  33
(1991).

\bibitem{BGS92}
C. Best, W. Greiner and G. Soff, Phys. Rev. {\bf A46} (1992) 261.

\bibitem{Vid93}
M. Vidovi\'{c}, M. Greiner, C. Best and G. Soff, Phys. Rev. {\bf
C47}, 2308 (1993).

\bibitem{Hen98}
K. Hencken, D. Trautmann and G. Baur, Phys. Rev. {\bf A51} (1995)
998; {\bf A51}, 1874 (1995).

\bibitem{Ba97}  A.J. Baltz, Phys. Rev. Lett. {\bf 78} (1997)
1231.

\bibitem{Ba96}  G. Baur {\it et al.}, Phys. Lett. {\bf B368},
351 (1996).

\bibitem{Bla80}
G. Blanford {\it et al.}, Phys. Rev. Lett. {\bf 80}, 3037 (1998).

\bibitem{Bec87}  U. Becker, J. Phys. {\bf B20}, 6563 (1987).

\bibitem{Mom91}
K. Momberger, N. Gr\"{u}n and W. Scheid, Z. Phys. {\bf D18}, 133
(1991).

\bibitem{Rum93}
K. Rumrich, G. Soff and W. Greiner, Phys. Rev. {\bf A47 },
 215 (1993).

\bibitem{Mom95}
K. Momberger, A. Belkacem and A.H. Sorensen, Europhys. Lett.{\bf \
32},  401 (1995).

\bibitem{Bal98}
A.J. Baltz and L. McLerran, Phys. Rev. {\bf C58}, 1679 (1998).

\bibitem{SW98}
 B. Segev and J.C. Wells, Phys. Rev. {\it A57},
1849 (1998).

\bibitem{Iva99}
D.Yu. Ivanov, A. Schiller, V.G. Serbo, Phys. Lett. {\bf B454},
155 (1999).

\bibitem{Eic99}
U. Eichmann, J. Reinhardt, S. Schramm and W. Greiner, Phys. Rev.
{\bf A59}, 1223 (1999).


\bibitem{LeM01}
R.N. Lee and A.J. Milstein, hep-ph/0103212.

\bibitem{BD00}  C. A. Bertulani and D. Dolci, Nucl. Phys. {\bf A683}, 635
(2001).

\bibitem{He00}  H. Meier, Z. Halabuka, K. Hencken, D. Trautmann and
G. Baur, e-print nucl-th/0008020.

\bibitem{BN02}
C.A. Bertulani and F. Navarra, Nucl. Phys. {\bf A703}, 861 (2002).

\bibitem{Low60}
F.E. Low, Phys. Rev. {\bf 120}, 582 (1960).

\bibitem{Ya50}
C.N. Yang, Phys. Rev. {\bf 77}, 242 (1950); L. Wolfenstein and
D.G. Ravenhall, Phys. Rev. {\bf 88}, 279 (1952).

\bibitem{BF90}
G. Baur and L.G. Ferreira Filho, Nucl. Phys. {\bf A518}, 786
(1990).

\bibitem{BYG85}
G.T. Bodwin, D.R. Yennie and M.A. Gregorio, Rev. Mod. Phys. {\bf
57}, 723 (1985).

\bibitem{Sap90}
``Theory of hydrogenic bound states'', J. Sapirstein and D.R.
Yennie, in ``Quantum electrodynamics'', edited by T. Kinoshita,
World Scientific, Singapore (1990).

\bibitem{Nov78} V.A. Novikov et al., Phys. Rep. {\bf 41C}, 1 (1978).

\bibitem{Pa89}
E. Papageorgiu, Phys. Rev. {\bf D40}, 92 (1989).

\bibitem{BLP82}
V. B. Berestetskii, E.M. Lifshitz and L.P. Pitaevskii, {\it
Quantum Electrodynamics}, 2nd edition, Pergamon, Oxford, 1982.

\bibitem{AP75}
T. Appelquist and H.D. Politzer, Phys. Rev. Lett. {\bf 34}, 43
(1975).

\bibitem{RW67}
R.P. Royen and V.F. Weisskopf, Nuovo Cim. {\bf A50}, 617 (1967).

\bibitem{Th70}
E.D. Theriot, Jr., R. H. Beers, V. W. Hughes, and K. O. H. Ziock,
Phys. Rev. {\bf A2}, 707 (1970).

\bibitem{Grab89}
M. Grabiak, B. M\"uller,  W. Greiner and P. Koch, J. Phys. {\bf
G15}, L25 (1989).

\bibitem{RN00}  C.G. Rold\~ao and A.A. Natale, Phys. Rev. {\bf
C61}, 064907 (2000).

\bibitem{Ja74}
C.W. de Jager, H. de Vries and C. de Vries, Atomic Data and
Nuclear Data Tables {\bf 14}, 479 (1974).

\bibitem{CJ90}
R.N. Cahn and J.D. Jackson, Phys. Rev. {\bf D42}, 3690 (1990).

\bibitem{VB02}
V.P. Gon\c calves and C.A. Bertulani, Phys. Rev. {\bf C65}, 054905
(2002).

\bibitem{sps}
See, e.g., U. Heinz, { Nuc. Phys. A } {\bf 685}, 414 (2000).

\bibitem{eskqm}
K. J. Eskola, Proceedings "Quark Matter 2001",  Nuc. Phys. A  (in
press), hep-ph/0104058.

\bibitem{vni}  See, e.g., K. Geiger., Phys. Rep. {\bf
258}, 237 (1995); X.-N Wang., Phys. Rep. {\bf 280}, 287 (1997).

\bibitem{caldwell}  H. Abramowicz and A. C. Caldwell, Rev. Mod. Phys.
{\bf 71},  1275 (1999).

\bibitem{mcl}
E. Iancu, A. Leonidov and L. McLerran, Nuc. Phys. {\bf A692}, 583
(2001).

\bibitem{mue}  A. H. Mueller,  Nucl. Phys.  {\bf B558}, 285 (1999).

\bibitem{vicsat}
M. B. Gay Ducati and V. P. Gon\c{c}alves, Phys. Lett.  {\bf B502},
92 (2001).

\bibitem{agl}
  A. L. Ayala, M. B. Gay Ducati and E. M. Levin. Nucl.
Phys. {\bf B493}, 305 (1997).

\bibitem{mclgos} E. Gotsman, E. Levin, U. Maor, L. McLerran and K. Tuchin, Nuc.
Phys.  {\bf A683}, 383 (2000).

\bibitem{df2a}  M. B. Gay Ducati and V. P. Gon\c {c}alves, Phys. Lett.
{\bf B466}, 375 (1999).

\bibitem{vicslope}
V. P. Gon\c{c}alves,  Phys. Lett. {\bf B495}, 303 (2000).

\bibitem{arneodo}
M. Arneodo, { Phys. Rep.} {\bf 240}, 301 (1994).

\bibitem{weise}
G. Piller and W. Weise, { Phys. Rep.} {\bf 330}, 1 (2000).

\bibitem{heraa}  M. Arneodo {\sl et al}., {\sl Future Physics at HERA.}
Proceedings of the Workshop 1995/1996. Edited by G. Ingelman {\sl
et al.}. hep-ph/9610423.

\bibitem{raju} R. Venugopalan, hep-ph/0102087.

\bibitem{perihe1}
Ch. Hofmann, G. Soff, A. Schafer and W. Greiner, { Phys. Lett.  }
{\bf B262}, 210 (1991).

\bibitem{perihe2}
N. Baron and G. Baur,  Phys. Rev.  {\bf C48}, 1999 (1993).

\bibitem{perihe3}
M. Greiner, M. Vidovic, Ch. Hofman, A. Schafer and G. Soff, Phys.
Rev. C {\bf C51}, 911 (1995).

\bibitem{perihe4}
F. Krauss, M. Greiner and G. Soff, {Prog. Part. Nucl. Phys.} {\bf
39}, 503 (1997).

\bibitem{grv95}  M. Gl\"{u}ck, E. Reya and A. Vogt,   Z. Phys. C {\bf
67}, 433 (1995).

\bibitem{eks}  K. J. Eskola, V. J. Kolhinen and C. A. Salgado,  Eur. Phys.
J. C {\bf 9 }, 61 (1999);   K.J. Eskola, V. J. Kolhinen and P. V.
Ruuskanen,  Nucl. Phys. B {\bf B535}, 351 (1998).

\bibitem{ag}
A.L. Ayala and  V. P. Gon\c{c}alves,  Eur. Phys. J.  {\bf C20},
343 (2001).

\bibitem{Gluck78}
M. Gl\"{u}ck and E. Reya, Phys. Lett.  {\bf 79}, 453 (1978).

\bibitem{nisius}
R. Nisius, Phys. Rep. {\bf 332}, 165 (2000).

\bibitem{frixione}
S. Frixione et al., Phys. Lett.  {\bf B348}, 633 (1995); Nucl.
Phys.  {\bf B454}, 3 (1995).

\bibitem{combridge}
B. L. Combridge, Nucl. Phys. {\bf B151}, 429 (1979).

\bibitem{grvphoton}
M. Gl\"{u}ck, E. Reya and A. Vogt, Phys. Rev. {\bf D45}, 3986
(1992)

\bibitem{gelis}
F. Gelis and A. Peshier, {Nuc. Phys. A}  (in press),
hep-ph/0107142.

\bibitem{revjpsi}
B. Naroska, hep-ex/0110023.

\bibitem{dl}
A. Donnachie and P. V. Landshoff, Nuc. Phys.  {\bf B244}, 322
(1984).

\bibitem{brod}  M. G. Ryskin, Z. Phys. {\bf C57}, 89 (1993); S. J.
Brodsky, L. Frankfurt, J. F. Gunion, A. H. Mueller and M.
Strikman, Phys. Rev. {\bf D50}, 3134 (1994); M. G. Ryskin, R. G.
Roberts, A. D. Martin and E. M. Levin,   Z. Phys.  {\bf C76}, 231
(1997); L. Frankfurt, W. Koepf and M. Strikman, Phys. Rev.  {\bf
D57}, 512 (1998).

\bibitem{fran}
L. Frankfurt, M. McDermott and M. Strikman, JHEP {\bf 03}, 045
(2001); K. Susuki, A. Hayashigaki, K. Itakura, J. Alam and T.
Hatsuda, Phys. Rev.  {\bf D62}, 031501 (2000).

\bibitem{BHT98}  G. Baur, K. Hencken and D. Trautmann, J. Phys. {\bf
G24}, 1657 (1998).

\bibitem{Dy77}  F. Dydak et al., Nucl. Phys. {\bf B118}, 1 (1977); P.C.
Petersen, A. Beretvas, T. Devlin, K. B. Luk, G. B. Thomson, and R.
Whitman, R. Handler, B. Lundberg, L. Pondrom, M. Sheaff, and C.
Wilkinson,  P. Border, J. Dworkin, O. E. Overseth, R. Rameika, and
G. Valenti, K. Heller and C. James, Phys. Rev. Lett., {\bf 57},
949 (1986).

\bibitem{Ant87}  Yu.M. Antipov, V. A. Batarin, V. A. Bessubov,
N. P. Budanov, Yu. P. Gorin, S. P. Denisov, S. V. Klimenko,
I. V. Kotov, A. A. Lebedev, A. I. Petrukhin, S. A. Polovnikov, and
D. A. Stoyanova , Phys. Rev. {\bf D36}, 21
(1987).

\bibitem{Ant83}  Yu.M. Antipov et al., Phys. Lett. {\bf B121} (1983) 445; Z.
Phys. {\bf C26}, 495 (1985).

\bibitem{spencer3}  S. Klein and E. Scannapieco, RHIC-STAR Note 1997/243,
"Two-photon physics with STAR"; G. Baur et al., LHC-CMS Note
2000/060, "Heavy ion physics programme in CMS"; H. Takai (for the
ATLAS experiment), private communication.

\end{thebibliography}
\end{document}